\documentclass[pre,notitlepage]{revtex4-1}
\usepackage{dcolumn}
\usepackage{wrapfig}
\usepackage{epsfig}
\usepackage{units}
\usepackage{dsfont}
\usepackage{comment}
\usepackage{amsfonts,amssymb,amsmath,bm}
\usepackage{graphicx,epsfig,color}

\usepackage{graphicx}
\usepackage{caption}
\usepackage{subcaption}

\newcommand\scalemath[2]{\scalebox{#1}{\mbox{\ensuremath{\displaystyle #2}}}}

\makeatletter
\begin{document}
\global\long\def\argmin{\operatornamewithlimits{argmin}}
\global\long\def\argmax{\operatornamewithlimits{argmax}}

\title{Ensemble of Thermostatically Controlled Loads: Statistical Physics Approach}
\author{Michael Chertkov$^{1,2}$, Vladimir Chernyak$^{3}$} 
\affiliation{$^{1}$Center for Nonlinear Studies \& T-4, Theoretical Division, Los Alamos National Laboratory, Los Alamos, NM 87545, USA}
\affiliation{$^{2}$Skolkovo Institute of Science and Technology, 143026 Moscow, Russia}
\affiliation{$^{3}$Department of Chemistry, Wayne State University, 5101 Cass Ave,Detroit, MI 48202, USA}

\maketitle

{\bf Thermostatically Controlled Loads (TCL), e.g. air-conditioners and heaters, are by far the most wide-spread consumers of electricity. Normally the devices are calibrated to provide the so-called bang-bang control of temperature -- changing from on to off, and vice versa, depending on temperature. Aggregation of a large group of similar devices into a statistical ensemble is considered, where the devices operate following the same dynamics subject to stochastic perturbations and randomized, Poisson on/off switching policy. We analyze, using theoretical and computational tools of statistical physics, how the ensemble relaxes to a stationary distribution and establish relation between the relaxation and statistics of the probability flux, associated with devices' cycling in the mixed (discrete, switch on/off, and continuous, temperature) phase space. This allowed us to derive and analyze spectrum of the non-equilibrium (detailed balance broken) statistical system and uncover how switching policy affects oscillatory trend and speed of the relaxation. Relaxation of the ensemble is of a practical interest because it describes how the ensemble recovers from significant perturbations, e.g. forceful temporary switching off aimed at utilizing flexibility of the ensemble in providing  "demand response" services relieving consumption temporarily to balance larger power grid. We discuss how the statistical analysis can guide further development of the emerging demand response technology.}

\begin{figure}[t]
\begin{center}
\includegraphics[scale=0.25,page=1]{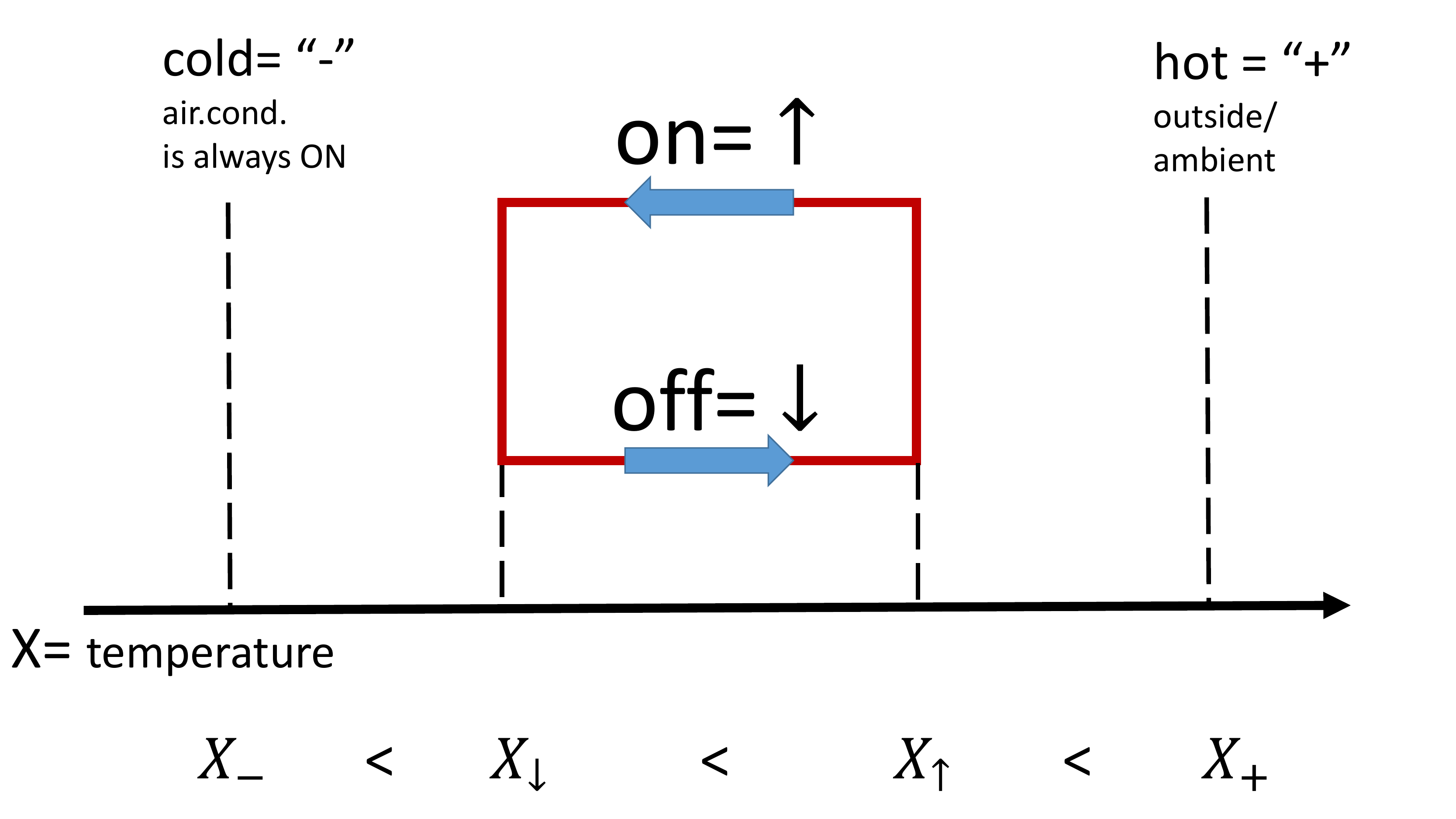}
\caption{{\bf Basic TCL (cooling) model.} }
\label{fig:on_off}
\end{center}
\end{figure}

Thermostatically Controlled Loads (TCL), or even more generally loads which cycle through multiple stages, came into focus of the engineering community, and specifically power engineering communities, in 80ies  and 90ies \cite{79CD,81IS,84CM,85MC,88MC}.  During the last decade the subject turned to become a central piece of the Demand Response (DR) paradigm \cite{04LC,05LCW,09Cal,11CH,11BF}.

In this paper we continue this trend and study TCL models, approaching it from the side of statistical/mathematical physics. Our basic TCL model, describing stochastic dynamics of a cooling device, e.g. air-conditioner (heating device, e.g. heater or boiler, can be described similarly), is stated  in terms of the following stochastic system of equations, introduced and discussed in \cite{79CD,81IS,85MC,88MC}, and illustrated in Fig.~(\ref{fig:on_off})
\begin{eqnarray}\label{eq:temp}
&& \frac{dx}{dt}=-\frac{1}{\tau}\left\{
\begin{array}{cc} x-x_+,& \sigma=\downarrow\\ x-x_-,\quad \sigma=\uparrow\end{array}\right. +\xi(t),\\
\label{eq:hard}
&& \sigma(t+dt)=
\left\{\begin{array}{cc}
\downarrow, & \sigma(t)=\uparrow \ \& \ x<x_\downarrow \\
\uparrow, &  \sigma(t)=\downarrow \ \& \ x>x_\uparrow \\
\sigma(t), & \mbox{otherwise}
\end{array}\right.
\end{eqnarray}
where the dynamic variables evolving in time, $t$, are continuous, $x(t)$, describing temperature, and binary, $\sigma(t)=\downarrow,\uparrow$, describing if the device is in the {\bf on} position or {\bf off} position, respectively. The deterministic part of the model (\ref{eq:temp},\ref{eq:hard}) is parameterized by the equilibrium temperatures, $x_-$ and $x_+$, settled if the device would be kept {\bf on}/{\bf off}, respectively, for the time longer than the relaxation time, $\tau$, and  the switching temperatures, $x_\downarrow$ and $x_\uparrow$, describing when the device which was {\bf on}/{\bf off} at $x>x_{\downarrow/\uparrow}$ switches {\bf off}/{\bf on} at $x=x_{\downarrow/\uparrow}$. The stochastic component, imitating effect of environment uncertainty (on temperature), is modeled by the zero mean short/white-correlated Gaussian Langevien term, $\xi(t)$, thus fully described by the covariance,
$\mathbb{E}[\xi(t_1)\xi(t_2)]=\kappa\delta(t_1-t_2)$, where $\kappa$ is the thermal drift/diffusion coefficient.

In what follows we call the model described by Eq.~(\ref{eq:temp},\ref{eq:hard}) the {\bf hard model}, to contrast it with the {\bf soft model} assuming probabilistic, thus soft, correction of the following type
\begin{eqnarray} &&
\mbox{If $\sigma(t+dt)\neq \sigma(t)$, accept $\sigma(t+dt)=\downarrow/\uparrow$  from Eq.~(\ref{eq:hard})}\nonumber\\
&& \mbox{with probability }r dt, \label{eq:soft}
\end{eqnarray}
where $r$ is the rate of the i.i.d. Poisson process.  The {\bf soft model} implements the idea of communication-limited control, where the aggregator communicates to all members of the ensemble only one number -- the Poisson rate $r$. Naturally,  hard model is recovered in the $r\to\infty$ limit of the soft model. (Generalization of Eq.~(\ref{eq:soft}) and following considerations allowing different Poisson rates depending on if the device is switched {\bf on} or {\bf off} is straightforward.) A model similar to our {\bf soft model} was discussed in \cite{12AK}.

In this manuscript we analyze the {\bf soft} model, as well as its {\bf hard} limit, by theoretical and computational tools of statistical physics. Main results reported in this paper are:
\begin{itemize}
\item[(1)] We derive analytic expression for a steady Probability Distribution Function (PDF) of the $(x,\sigma)$ mixed state within the {\bf hard model} and relate it to the flux of probability in this non-equilibrium setting where the detailed-balance condition is broken.
\item[(2)] We analyze Fokker-Planck (FP) equation for the PDF of the $(x,\sigma)$ state in a non-steady, transient regime, representing solution through the spectral expansion over the (right) eigen-values of the FP operator:
    \begin{eqnarray}
    P_\sigma(x|t)=\sum_{n}\exp(-\lambda_n t) \xi_{n}(x,\sigma),\label{eq:spectral}
    \end{eqnarray}
    where $\mbox{Re}(\lambda_{n+1})>\mbox{Re}(\lambda_n)>\cdots >\mbox{Re}(\lambda_0)=0$. Then we express $\xi_\sigma(x,\sigma)$ explicitly in terms of the hyper-geometric functions and present $\lambda_n$ implicitly, as a solution of a system of ($8$ in the case of soft model and $4$ in the case of hard model) transcendental equations. We analyze the system of equations numerically in the case of the hard model, e.g. showing that $\lambda_n$ for $n\leq 1$ contains both real (decaying) and imaginary (oscillatory) parts.
\item[(3)] We study the {\bf soft} model in the diffusionless limit of $\kappa\to 0$ and establish that the system mixes, i.e. a steady state is achieved, with a mixing rate exponential in time. We derive explicit transcendental equation describing spectrum. This allows theoretical and an efficient computational analysis of the spectrum, controlling relaxation. One observes that, if the switching rate is sufficiently small, $r\tau\ll 1$, oscillations of the probability distribution (as it transitions to the steady state) are suppressed, and the relaxation is split in two stages, fast but independent equilibration within the on and off states, which occurs with the rate $2/\tau$, followed by a slower and non-equilibrium mixing between the on/off states occurring with the switching rate, $r$. In the opposite regime of the large switching rate, $r\tau\gg 1$, one observes persistent oscillations with the period
    \begin{eqnarray}
    t_{dc}=\tau \log\left(\frac{(x_\uparrow-x_-)(x_+-x_\downarrow)}{(x_\downarrow-x_-)(x_+-x_\uparrow)}\right),
    \label{t_dc}
    \end{eqnarray}
    associated with the deterministic (phase space) cycling, eventually decaying with the mixing rate equal to the switching rate, $r$. We also discuss (e.g. briefly) effect of a small, but finite, diffusion.
\item[(4)] For the {\bf soft} model in the dimensionless regime we  establish explicit relation between dynamics of the PDF of the mixed $(x,\sigma)$ state and finite time PDF of the flux, defined as the number of cycles made in the phase space in time. The latter object measures degree of  how non-equilibrium (far from detailed balance) the system is.  This relation and analysis are methodologically important as being, to the best of our knowledge, the first of this kind among other models of the non-equilibrium statistical physics.
    \end{itemize}

Organization of material in the remainer of the paper is as follows. Derivation and properties of the basic equations governing temporal evolution of the Probability Distribution Function (PDF) of $x,\sigma$, the FP equation, are discussed in Section \ref{sec:FP}. Section \ref{sec:analysis} is devoted to analysis of (a) stationary solution of FP equation and (b) spectral analysis of the {\bf hard} model.  Soft model is analyzed in Section \ref{sec:soft} consisting of three Subsections discussing, e.g., the cases of zero diffusivity and small diffusivity.
Results of the direct numerical simulations of the FP equations, validating and confirming consistency of the aforementioned analysis are presented in Section \ref{sec:numerics}. Section \ref{sec:Lagr_dyn} is devoted to presenting Lagrangian, i.e. a single device moving through the phase space, prospectives on the systems dynamics. Section \ref{sec:conclusions} is reserved for conclusions and discussion of the path forward.

\section{Fokker-Planck Equations}
\label{sec:FP}

Next we discuss temporal dynamics of the PDF of the, $(x,\sigma)$, state, $P_\sigma(x|t)$, where $\sigma=\uparrow,\downarrow$. The so-called Fokker-Planck equations, governing dynamics of $P_\sigma(x|t)$, follow straightforwardly from the stochastic differential Eqs.~(\ref{eq:temp},\ref{eq:hard},\ref{eq:soft}). (See e.g. classic statistical physics textbooks, e.g. \cite{Feynman,vanKampen,Gardiner}, for general description of the derivation methodology.) FP equations for the {\bf soft model} becomes
\begin{eqnarray}
\label{FP-two-state_1}
&& \frac{\partial{P}_{\uparrow}}{\partial t}= \kappa\frac{\partial^{2}P_{\uparrow}}{\partial x^{2}}+ \frac{\partial}{\partial x}\left(\frac{x-x_-}{\tau}P_{\uparrow}\right)- r_{\downarrow\uparrow}(x)P_{\uparrow}+ r_{\uparrow\downarrow}(x)P_{\downarrow}, \\
\label{FP-two-state_2} &&
\frac{\partial{P}_{\downarrow}}{\partial t}= \kappa\frac{\partial^{2}P_{\downarrow}}{\partial x^{2}}+ \frac{\partial}{\partial x}\left(\frac{x-x_+}{\tau}P_{\downarrow}\right)- r_{\uparrow\downarrow}(x)P_{\downarrow}+ r_{\downarrow\uparrow}(x)P_{\uparrow},\\ \label{rs}
&& r_{\uparrow\downarrow}(x)\doteq r\left\{\begin{array}{cc} 1,& x<x_\downarrow\\ 0,& x>x_\downarrow\end{array}\right.,
\quad r_{\downarrow\uparrow}(x)\doteq r\left\{\begin{array}{cc} 1,& x>x_\uparrow\\ 0,& x<x_\uparrow\end{array}\right. .
\end{eqnarray}
The equations should be supplemented by the natural zero boundary conditions and the normalization condition, respectively:
\begin{eqnarray}
&&x\to\pm\infty:\quad P_{\uparrow/\downarrow}(\pm\infty)=0, \label{bound_inf}\\
&&\int_{-\infty}^{+\infty} dx \left(P_\uparrow(t,x)+P_\downarrow(t,x)\right)=1.\label{normalization}
\end{eqnarray}
In the {\bf hard model} limit, i.e. at $r\to\infty$, the FP Eqs.~(\ref{FP-two-state_1},\ref{FP-two-state_2}) turn into
\begin{eqnarray}
\label{FP-hard_1}
&& \frac{\partial{P}_{\uparrow}}{\partial t}= \kappa\frac{\partial^{2}P_{\uparrow}}{\partial x^{2}}+ \frac{\partial}{\partial x}(\frac{x-x_-}{\tau}P_{\uparrow})+J_{\uparrow\downarrow}\delta(x-x_\uparrow), \\
\label{FP-hard_2} &&
\frac{\partial{P}_{\downarrow}}{\partial t}= \kappa\frac{\partial^{2}P_{\downarrow}}{\partial x^{2}}+ \frac{\partial}{\partial x}(\frac{x-x_+}{\tau}P_{\downarrow})+J_{\downarrow\uparrow}\delta(x-x_\downarrow),
\end{eqnarray}
supplemented by the so-called absorbing boundary conditions
\begin{eqnarray}
\label{FP-hard-BC} P_{\uparrow}(x_\downarrow)=0,\quad P_{\downarrow}(x_\uparrow)=0,
\end{eqnarray}
i.e. conditions reenforcing termination of a "particle"/airconditioner  after it reaches respective switching boundary. The ``switching-induced'' fluxes in Eqs.~(\ref{FP-two-state_1},\ref{FP-two-state_2}) are defined according to
\begin{eqnarray}
\label{FP-switching-fluxes} && J_{\uparrow\downarrow}= -\kappa \left(\frac{\partial P_{\downarrow}}{\partial x}\right)_{x=x_\uparrow}= -\kappa \partial_{x}P_{\downarrow}(x_\uparrow),\\
&& J_{\downarrow\uparrow}= \kappa \left(\frac{\partial P_{\uparrow}}{\partial x}\right)_{x=x_\downarrow}= \kappa \partial_{x}P_{\uparrow}(x_\downarrow).\nonumber
\end{eqnarray}
The "instantaneous" switch terms on the right hand sides of Eqs.~(\ref{FP-hard_1},\ref{FP-hard_2}), and (\ref{FP-switching-fluxes}) reflects the absorbing nature of the jumping process, i.e., immediate jump/switch once the hard boundary for respective $\uparrow,\downarrow$ state is achieved, and then Eq.~(\ref{FP-switching-fluxes}) reenforces relations between switching rates and respective probability fluxes at the hard boundaries. Note that the current fluxes at the r.h.s. of Eq.~(\ref{FP-switching-fluxes}) do not include the ``advection" induced ($\tau$-dependent) terms, since the latter vanish due to the boundary condition (\ref{FP-switching-fluxes}). Notice, that a system of equations for the {\bf hard model} equivalent to Eqs.~(\ref{FP-hard_1},\ref{FP-hard_2},\ref{FP-hard-BC},\ref{FP-switching-fluxes}) was first time derived in \cite{85MC}.

\section{Hard model: Steady Distribution \& Spectral Analysis}

In this Section, devoted to the hard model, we, first, derive stationary distribution and then analyze spectrum of the respective FP operator.

\subsubsection{Stationary Distribution}

In the stationary state of the hard model the fluxes of probability in the $\uparrow$ and $\downarrow$ states are piece-wise constant
\begin{eqnarray}\label{J-hard}
-\kappa\frac{\partial P_{\uparrow/\downarrow}}{\partial x}+\frac{x-x_{-/+}}{\tau}P_{\uparrow/\downarrow}=\left\{
\begin{array}{cc} \mp J, & x_\downarrow\leq x\leq\uparrow  \\
0, & \mbox{otherwise}\end{array}\right.
\end{eqnarray}
where $J$ is a positive constant, while $P_\uparrow(x)$ at $x\geq x_\uparrow$ and $P_\downarrow(x)$ at $x\leq x_{\downarrow}$ admit, in accordance with Eqs.~(\ref{bound_inf}) the following equilibrium (Boltzmann) forms, $\sim \exp(-k(x- x_-)^{2}/(2\kappa))$.
Then, solving Eqs.~(\ref{J-hard}) and accounting for the normalization condition (\ref{normalization}) one arrives at the following explicit expressions
\begin{eqnarray} \label{rho_g_stat}
&& P_{\uparrow/\downarrow} (x)=\frac{J e^{-\frac{(x-x_\mp)^2}{2\kappa\tau}}}{\kappa}
g_{\uparrow/\downarrow} (x),\\ && \label{g_up_stat}
g_{\uparrow}(x)
\doteq
\left\{\begin{array}{cc}
0,& x\leq x_\downarrow\\
\int\limits_{x_\downarrow}^x dx' e^{-\frac{(x'-x_-)^2}{2\kappa\tau}},&
x_\downarrow\leq x\leq x_\uparrow \\
\int\limits_{x_\downarrow}^{x_\uparrow} dx' e^{-\frac{(x'-x_-)^2}{2\kappa\tau}},&
x_\uparrow\leq x\end{array}\right.
\\ && \label{g_down_stat}
g_\downarrow (x)=
\left\{\begin{array}{cc}
0,& x\geq x_\uparrow\\
\int\limits_x^{x_\uparrow} dx' e^{-\frac{(x'-x_+)^2}{2\kappa\tau}},&
x_\downarrow\leq x\leq x_\uparrow\\
\int\limits_{x_\downarrow}^{x_\uparrow} dx' e^{-\frac{(x'-x_-)^2}{2\kappa\tau}},&
x_\downarrow\leq x\leq x_\uparrow
\end{array}\right.\\ \label{J_stat}
&& J\doteq \frac{\kappa}{\int\limits_{-\infty}^{+\infty} dx \left(g_\uparrow(x) e^{-\frac{(x-x_-)^2}{2\kappa\tau}}+g_\downarrow(x) e^{-\frac{(x-x_+)^2}{2\kappa\tau}}\right)}.
\end{eqnarray}

\subsubsection{Spectral Analysis}

\begin{figure}[t]
\begin{center}
\includegraphics[scale=0.25]{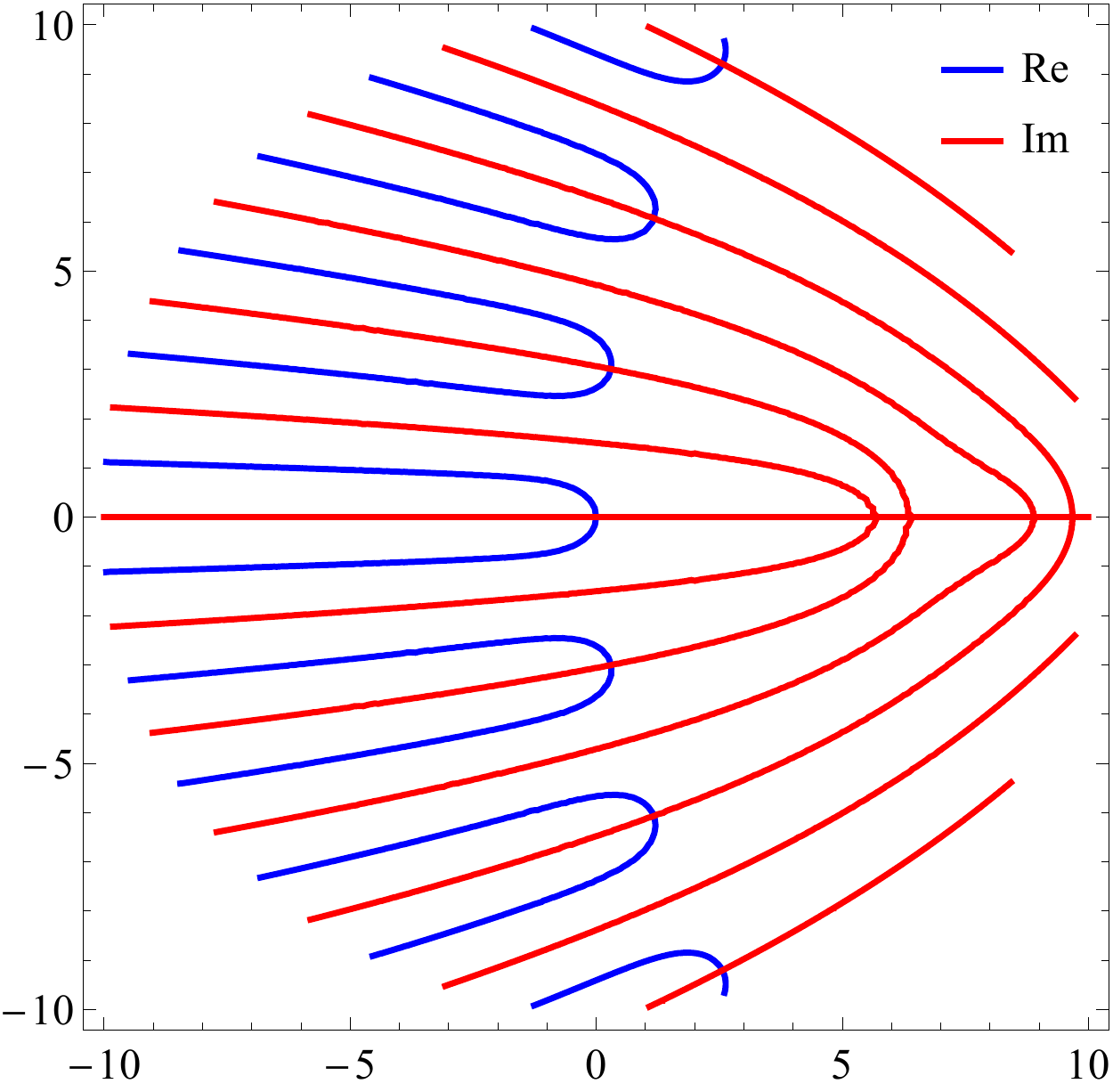}
\caption{{\bf Spectral Analysis of the Hard Model.} $\kappa=0.1$, $\tau=1$, $x_-=-2, x_\downarrow=-1, x_\uparrow=1, x_+=2$. The axis are $\mbox{Re}[\lambda_n]$ and $\mbox{Im}[\lambda_n]$. Red and blue lines mark isolines of the real and imaginary parts of $\mbox{det}(M(\lambda))$ respectively, see Appendix \ref{app:Hard-Spectral} for details. Any crossing of a blue line and a red line corresponds to an eignevalue, $\lambda_n$.}
\label{fig:hard-det}
\end{center}
\end{figure}

Solution of Eqs.~(\ref{FP-hard_1},\ref{FP-hard_2}) can be presented in the form of the eigen-function expansion for the left (ket) modes of the respective Fokker-Planck operator \cite{09Cal}
\begin{eqnarray}
\left(
\begin{array}{c}
P_\uparrow( x |t) \\ P_\downarrow( x |t)
\end{array}
\right)=\sum_n \exp\left(-t\lambda_n t\right)\left(
\begin{array}{c}
\xi_{\uparrow,n}(x) \\ \xi_{\downarrow,n}(x)
\end{array}
\right),
\label{eigen-exp}
\end{eqnarray}
where explicit expressions for left (ket) modes, $\xi_{\sigma,n}(x)$ (defined up to re-scaling by a convolution of the right (bra) modes with the initial condition) in terms of Kummer's confluent hypergeometric function and some further details are presented in the Appendix \ref{app:Hard-Spectral}; and it is assumed that . Positivity of the FP operator guarantees that the spectrum is discrete, and the eigenvalue with the lowest real part is zero. It is natural to order the eigenvalues accordingly, $\lambda_0=0\leq \mbox{Re}(\lambda_1)\leq \cdots \leq \mbox{Re}(\lambda_n)$, where $n$ is nonnegative integer.
Notice that, contrary to what was claimed in \cite{09Cal}, all nonzero eigenvalues are complex,  i.e. having nonzero real and imaginary parts.   Eigenvalues satisfy a system of transcendental equations stated in Appendix \ref{app:Hard-Spectral} explicitly as a zero condition for a determinant of a $4\times 4$ complex valued matrix. Numerical solution of the system of equations is illustrated in Fig.~(\ref{fig:hard-det}), showing iso-lines of real and imaginary parts of the determinant, where therefore crossing of the two identify four eigenvalues with the smallest real parts.

\section{Soft Model: Spectral Analysis}
\label{sec:soft}

In this Section we analyze spectrum of the Fokker-Planck operator of the soft model.  The analysis is done in two steps. First, we ignore diffusion completely and show that the transcendental equation for the spectral parameter outputs discrete spectrum. Then we discuss corrections acquired in the spectral equation due to small but finite diffusivity.

\subsection{Zero and Small Diffusivity}
\label{sub:soft_zero_dif}

We consider two cases different in view of the order the limit of zero diffusivity and infinite observation time are taken. The diffusivity will be considered small in both cases,  however in the first, $\kappa=0$, case we assume that first $\kappa\to 0$ and then the observation time, $t$, will be send to $\infty$. On the contrary in the case of the small $\kappa$,  the $t\to\infty$ limit is taken first, and only after that, $\kappa\to 0$.

\subsubsection{Zero diffusivity}
\label{subsub:zero_diff}

\begin{figure}[t]
\centering
\includegraphics[width=0.19\textwidth]{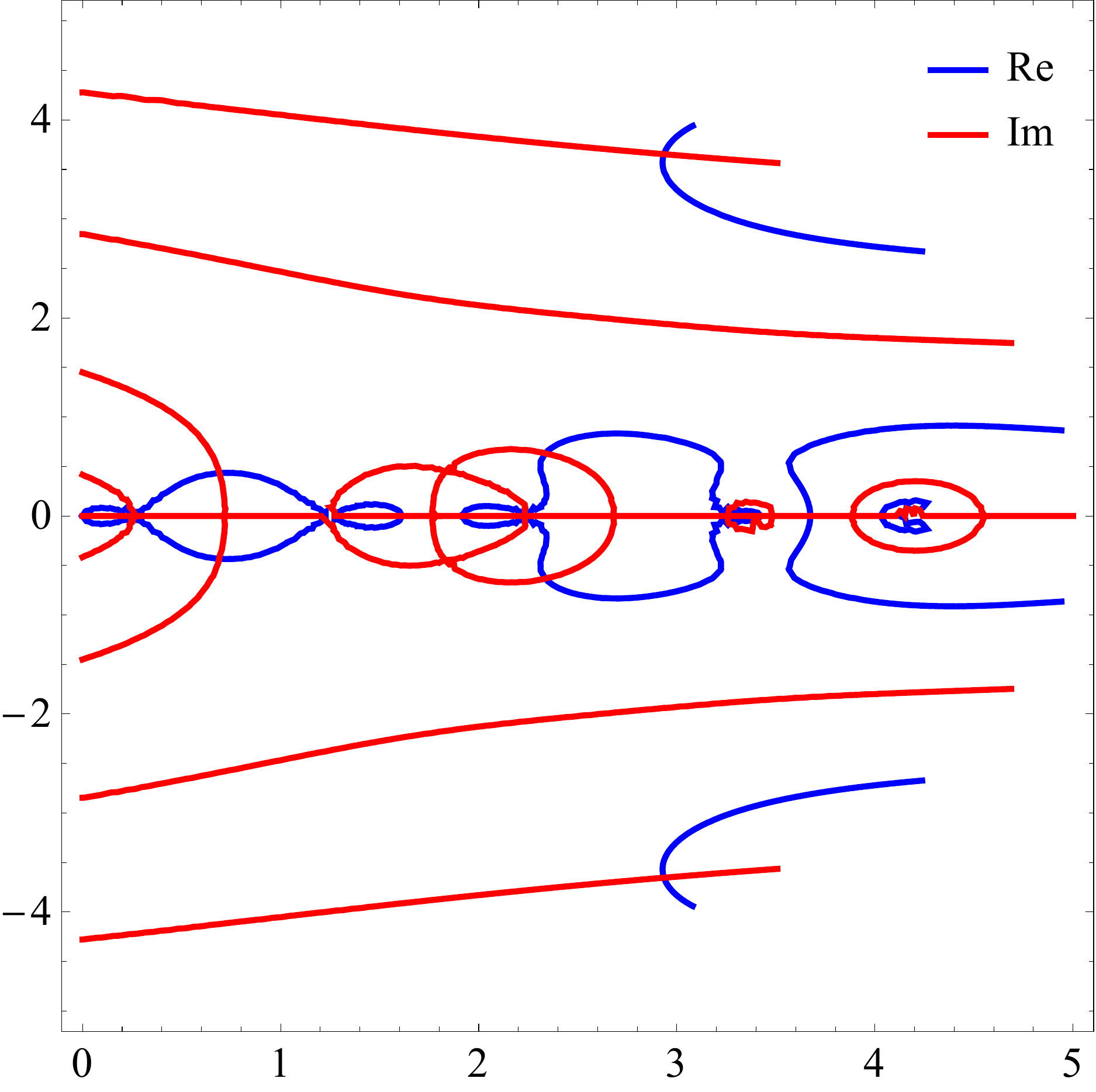}
\includegraphics[width=0.19\textwidth]{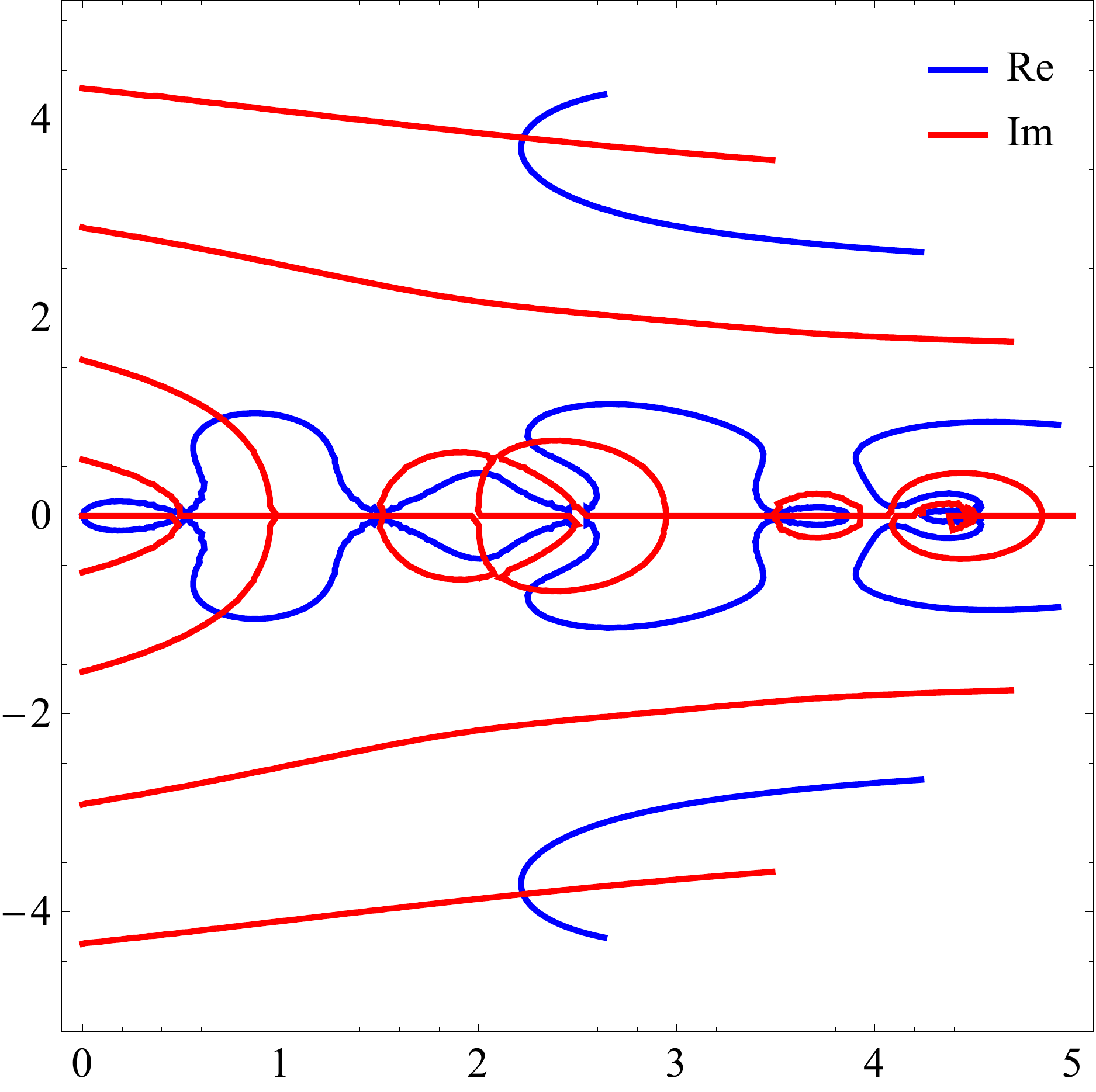}
\includegraphics[width=0.19\textwidth]{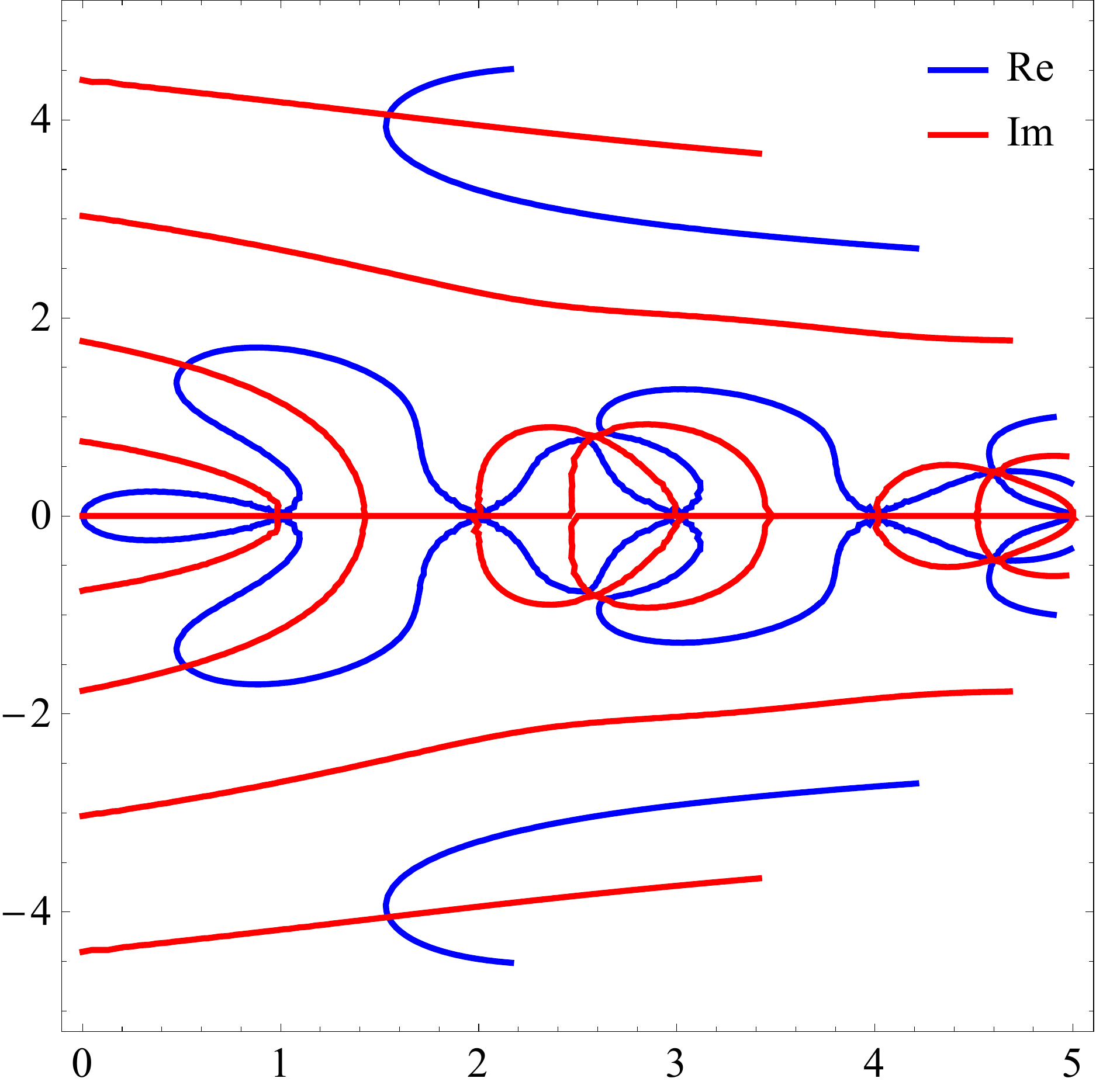}
\includegraphics[width=0.19\textwidth]{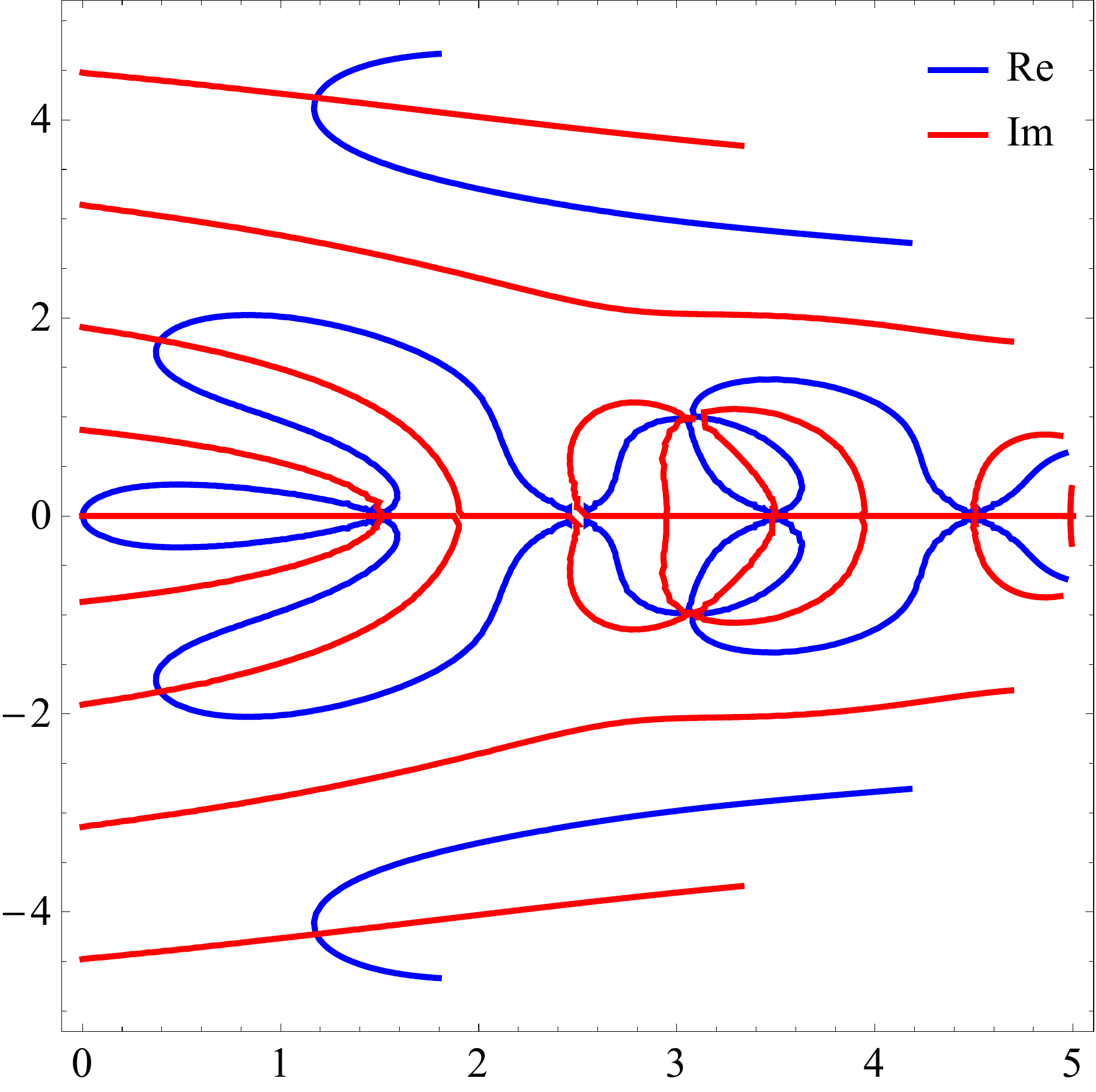}
\includegraphics[width=0.19\textwidth]{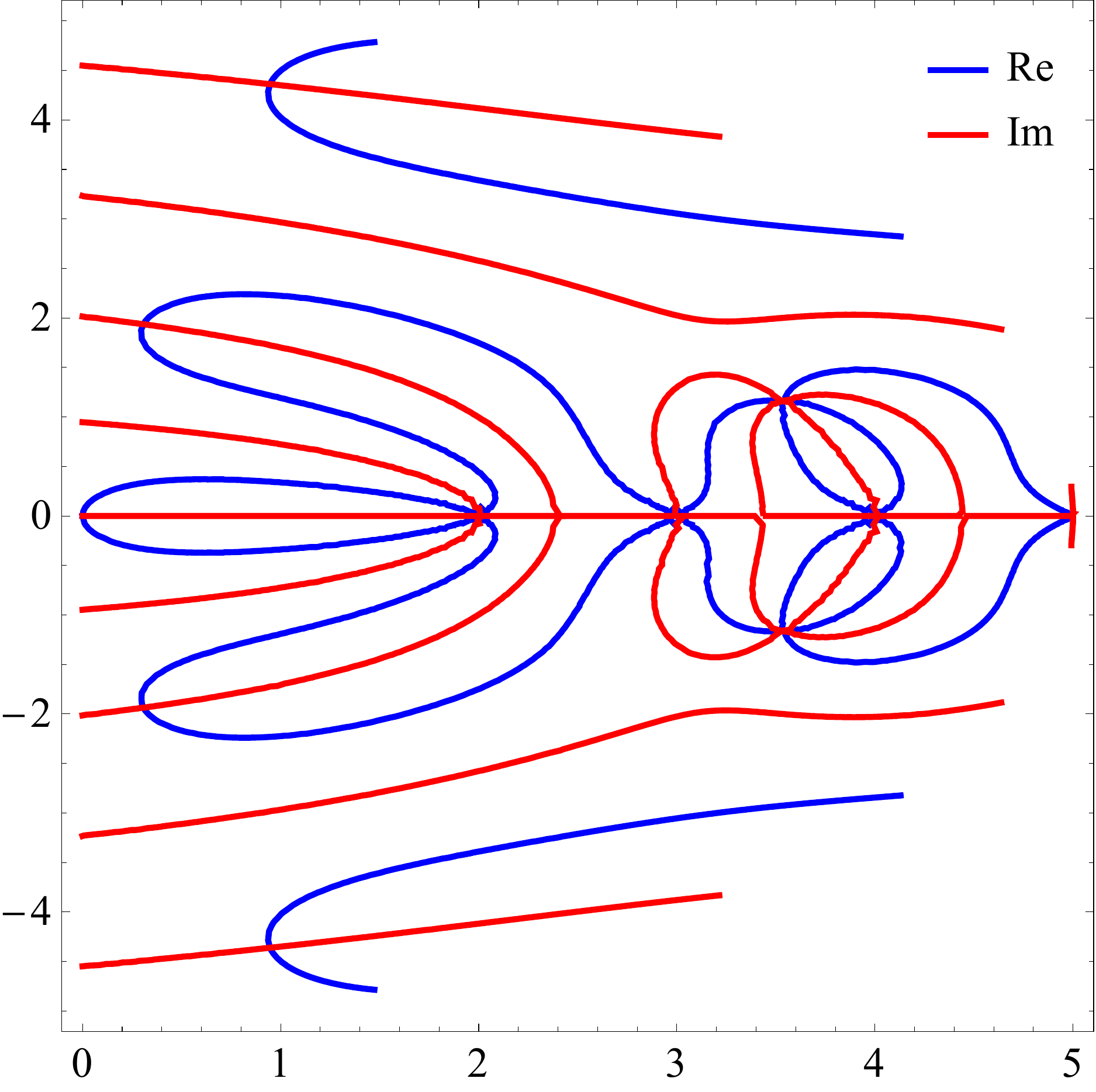}
\caption{Graphical solution of Eqs.~(\ref{zero-diff-spectral}) illustrated for $x_-=-2, x_\downarrow=-1, x_\uparrow=1,x_+=2$, $\tau=1$ and $r=0.5,1,1.5, 2$ in five sub-figures ordered, respectively, from left to right.
\label{fig:zero-diff-figs}}
\end{figure}

If the Langevin/diffusion term is ignored  Eqs.~(\ref{FP-two-state_1},\ref{FP-two-state_2}) combined with Eqs.~(\ref{eigen-exp}) transition to
\begin{eqnarray}
\label{FP-two-state-n-zero-diffusivity}
\left(\begin{array}{cc}
\lambda_n+\partial_x \frac{x-x_-}{\tau}-r_{\downarrow\uparrow}(x) & r_{\uparrow\downarrow}(x) \\
r_{\downarrow\uparrow}(x) & \lambda_n+\partial_x \frac{x-x_+}{\tau}-r_{\uparrow\downarrow}(x)
\end{array}\right)
\left(\begin{array}{c}\xi_{1,n}(x) \\ \xi_{2,n}(x)\end{array}\right)=0
\end{eqnarray}

Stability/spectral analysis of the noiseless model (\ref{FP-two-state-n-zero-diffusivity}), also focused mainly on discussion of heterogeneous ensembles the control design, was reported in \cite{15GMK}. In the following we present detailed analysis of the model.

Notice that $x\leq x_-$ and $x\geq x_+$ are not reachable in the Langevien/diffusion free regime.
In other (left, middle  and right) domains we can write down solutions explicitly up to constants (six) to be determined in the result of imposing proper boundary conditions. Consider,  first the left (-) interval. Here we write
\begin{eqnarray}
& x_-<x\leq x_\downarrow: &
\xi_{1,n}(x)= c_{1,-,n} (x-x_-)^{-1+(r-\lambda_n)\tau}
\label{zero-diff-left_1}\\
&& \xi_{2,n}(x)=c_{1,-,n}r\tau(x_+-x)^{-1-\lambda_n\tau}\int\limits_{x_-}^x dx' \frac{(x_+-x')^{\lambda_n\tau}}{(x'-x_-)^{1+(\lambda_n-r)\tau}}
\label{zero-diff-left_2}
\end{eqnarray}
where we have accounted for the fact that devices cannot reach $x=x_-$ in the switched-off state (one of the six boundary conditions).
Respective expressions for the right interval are
\begin{eqnarray}
& x_\uparrow<x\leq x_+: &
\xi_{2,n}(x)= c_{2,+,n} (x_+-x)^{-1+(r-\lambda_n)\tau},
\label{zero-diff-right_2}\\
&& \xi_{1,n}(x)=c_{2,+,n}r\tau(x-x_-)^{-1-\lambda_n\tau}\int\limits_x^{x_+} dx' \frac{(x'-x_-)^{\lambda_n\tau}}{(x_+-x')^{1+(\lambda_n-r)\tau}}.
\label{zero-diff-right_1}
\end{eqnarray}
And solutions in the middle interval are simply
\begin{eqnarray}
& x_\downarrow\leq x\leq x_\uparrow: &
\xi_{1,n}(x)=c_{1,n} (x-x_-)^{-1-\lambda_n\tau},\quad \xi_{2,n}(x)=c_{2,n} (x_+-x)^{-1-\lambda_n\tau}.
\label{zero-diff-middle}
\end{eqnarray}
Relating the left, right and middle solution through the continuity requirement at $x_\downarrow$ and $x_\uparrow$ respectively one arrives at the following spectral condition
\begin{eqnarray}
\left(\int\limits_{x_-}^{x_\downarrow} dx \frac{(x_+-x)^{\lambda_n\tau}}{(x-x_-)^{1+(\lambda_n-r)\tau}}\right)\left(\int\limits_{x_\uparrow}^{x_+} dx \frac{(x-x_-)^{\lambda_n\tau}}{(x_+-x)^{1+(\lambda_n-r)\tau}}\right)=\left(r \tau\right)^2\left((x_\downarrow-x_-)(x_+-x_\uparrow)\right)^{r\tau}.
\label{zero-diff-spectral}
\end{eqnarray}
Direct check shows that $\lambda=0$ is a solution of Eq.~(\ref{zero-diff-spectral}) (as required by existence of the steady state). In general,  the spectrum is complicated as shown in a set of illustrative Figures (\ref{fig:zero-diff-figs}).

Of a special interest is the issue of convergence of the integrals, entering Eq.~(\ref{zero-diff-spectral}). Formally the integrals are convergent only for $\mbox{Re}(\lambda_n) < r$. However, the integrals allow efficient analytic continuation beyond the condition. In fact, to get illustrative/numerical results shown in Figs.~(\ref{fig:zero-diff-figs}) we first express the integrals in the region of their convergence via the hypergeometric functions, and then use Mathematica ability to use known analytical properties of the hypergeometric functions to analytically continue the results beyond the constraints.
In fact,  it is clearly seen that $\lambda_n=r$ itself shows prominently as a valid solution of Eq.~(\ref{zero-diff-spectral}).

Analyzing numerical experiments we observe that, as expected only solution with non-negative $\mbox{Re}(\lambda_n)$ are realized.  The spectrum seen in the simulations is rich. The following features are observed. In general solutions are complex, i.e. contain nonzero real and imaginary part.  The real part of the eigenvalue is always positive. $\lambda=0$, correspondent to the stationary solution is separated by a gap from the rest of the spectrum.  Eigenvalues with $\lambda_n\geq r$ are always real and the $\lambda+n=r$ solution is always present. When $r$ is sufficiently large an infinite sequence of solutions with
$\mbox{Re}(\lambda_n) <r$ and imaginary part increasing with $n$ (by absolute value) emerges. Thus, in this regime the long-time asymptotic is an oscillatory decay. However, when $r$ is sufficiently small the eigenvalue with the lowest real part is the aforementioned special one, $\lambda_n=r$, which is real - thus resulting in purely decaying long-time asymptotic. 

\subsubsection{Small diffusivity}
\label{subsub:small_diff}

The limit $\kappa \to 0$ of small (but finite) diffusivity is important due to the fact that noise/diffusion is always present in the system. The formal side of this statement is linked to the fact that the FP operator changes  in transition from the duffusionless case, when it is the first-order (hyperbolic), to the case of a small diffusion when the FP operator becomes of the second-order (elliptic).  This change in the operator order has principal consequences for the shapes of the stationary solution, spectrum and, respectively, dynamic behavior in certain regimes.  The goal of this Subsection is to discuss these principal changes briefly and informally, leaving more detailed mathematical analysis for future publications.

We start the discussion by mentioning that it is clear from analysis of the steady distribution conducted for the diffusionless regime above that accounting for small but finite diffusion will be important mainly at $r\tau<1$. Indeed, in this regime the PDFs of the {\bf on}/{\bf off} states are picked in a vicinity of $x_{\pm}$. On the other hand when the diffusion is strictly zero the $x_{\pm}$ limits themselves are not achievable.  Therefore, immediate vicinities of $x_{\pm}$ are controlled solely by the diffusion.

Two temporal scales significant in this regime are, first,  $1/r$,  which is the typical time of switching from {\bf on} to {\bf off} and vice versa,  and then, second, the time for the system to stay within the {\bf on} or {\bf off} states.  Notice that the later time is not just $\tau$, as one can naively suggest. Indeed, given that majority of devices in the $r\tau <1$ regime get rather close to $x_\pm$, while the dynamics in the vicinity of the equilibrium points slow down, the respective time for a device to traverse the range and then get inside the diffusion-controlled zone on the other end of the range is estimated as $\tau \log (|x_+-x_|/\sqrt{\kappa \tau})$. This suggests that the dynamics, spectrum and structure of eigenvalues discussed above in Section \ref{subsub:zero_diff} for the diffusionless regime,  is in fact realized within the following intermediate time/transient asymptotic, $t\ll \tau \log (|x_{+}-x_{-}|/\sqrt{\kappa \tau})$.

Let us know discuss the long time asymptotic regime, $t \gg \tau \log (|x_{+}-x_{-}|/\sqrt{\kappa \tau})$, dominated by the diffusion. First, we notice that
the stationary distribution, achieved in the result of a balance of the ballistic motion within a state and jumps between the states, is renormalized by small diffusion in the $|x-x_\pm|\sim \sqrt{\kappa r}$ vicinity of $x_\pm$. Similar renormalization of the $x$-shape by diffusion also applies to the rest of the spectrum (beyond the stationary solution).  In fact,  the spectrum itself in this long time regime is modified as well. In particular,  if $r$ is sufficiently small, i.e. $\tau r \log (|x_+-x_|/\sqrt{\kappa \tau})\ll 1$, one observes perfect equidistant spectrum $\lambda_n= n/\tau$, where $n=0,1,\cdots$, correspondent to separate equilibrations within the $\sigma=\uparrow$ and $\sigma=\downarrow$ ensembles.

\section{Numerical Experiments}
\label{sec:numerics}

\begin{figure}[t]
\centering
\begin{subfigure}{0.3\textwidth}
\centering
\includegraphics[width=\textwidth]{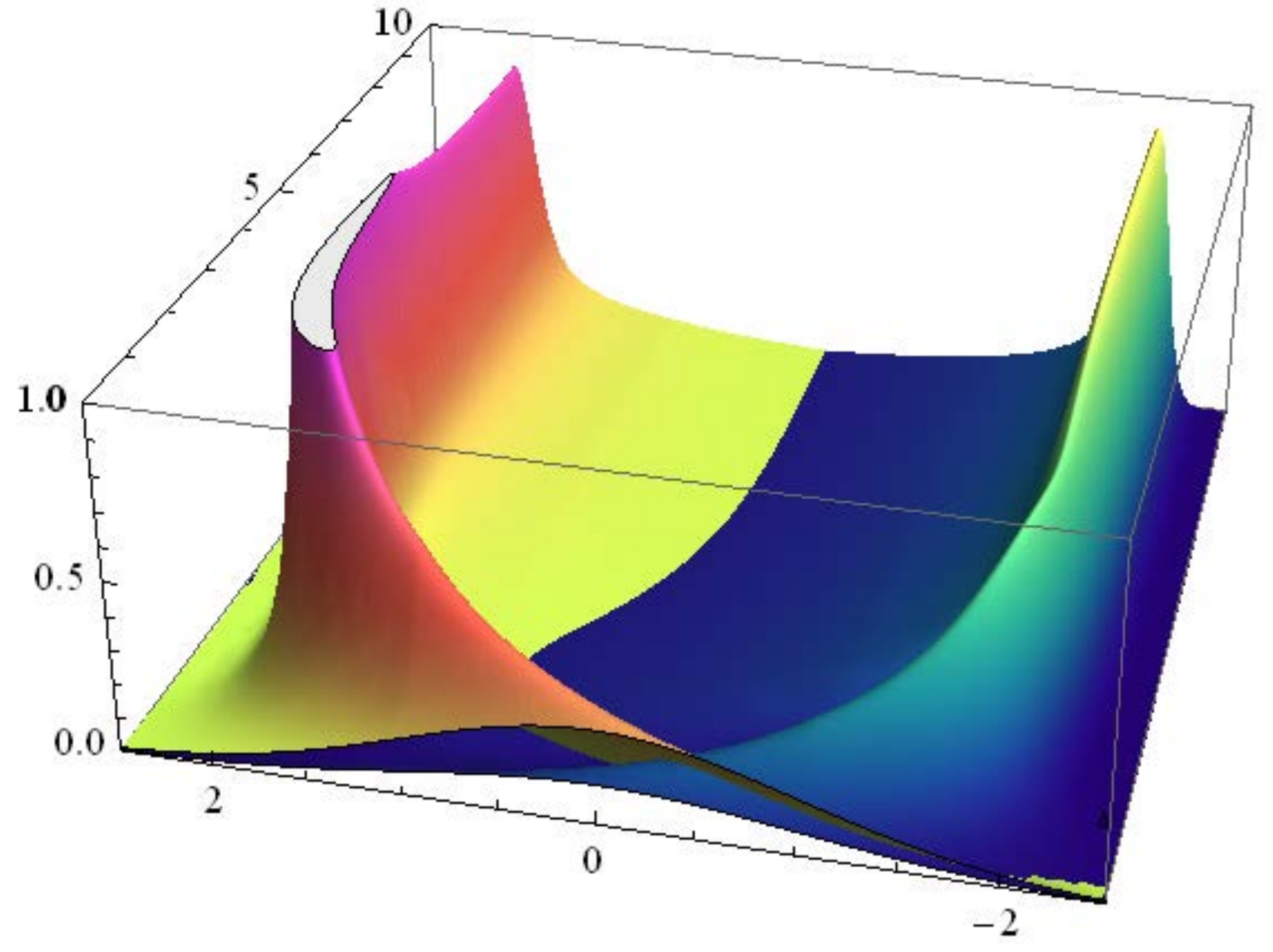}
\caption{$r=0.25$.}
\end{subfigure}
\begin{subfigure}{0.3\textwidth}
\centering
\includegraphics[width=\textwidth]{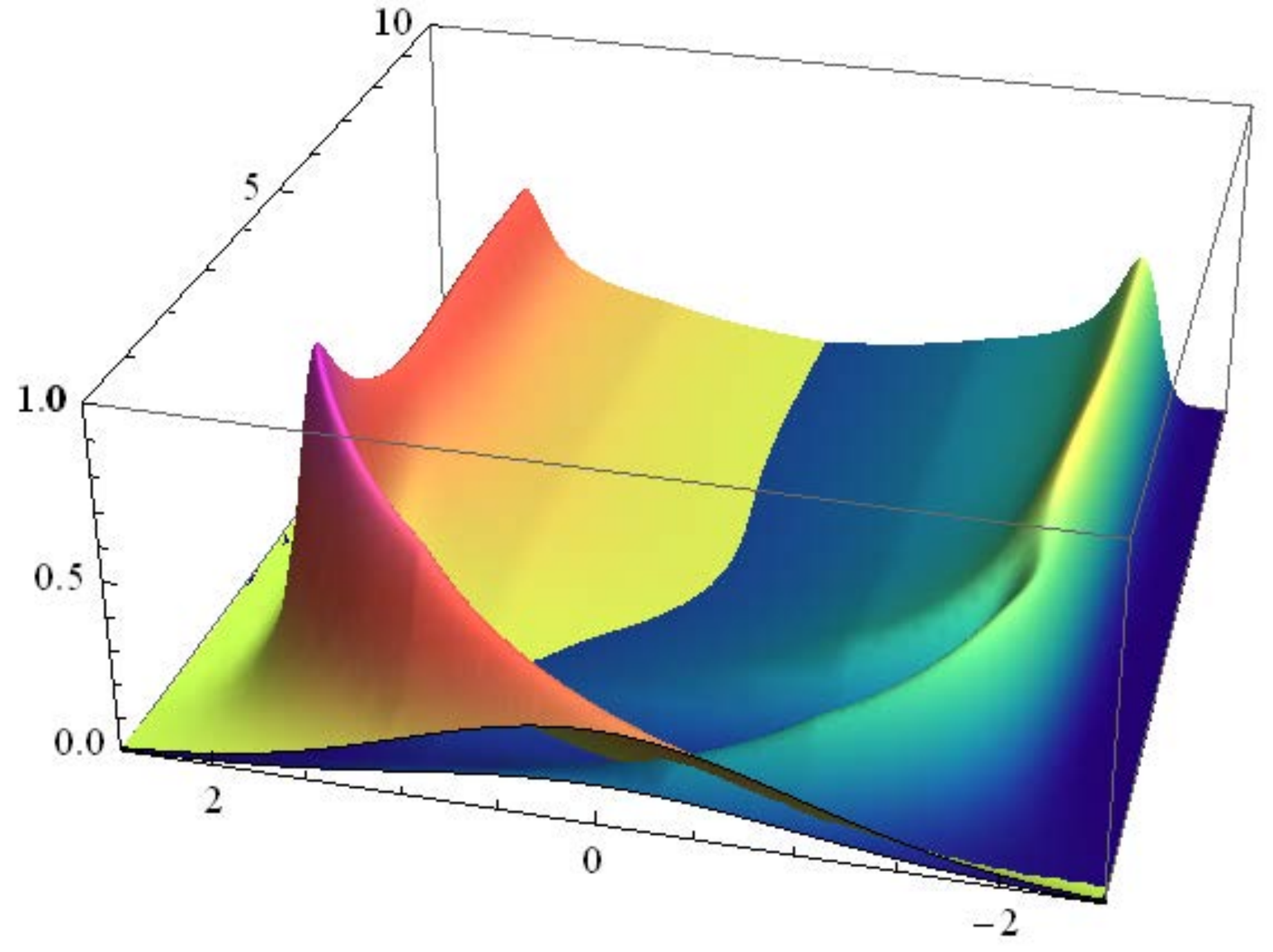}
\caption{$r=0.5$.}
\end{subfigure}
\begin{subfigure}{0.3\textwidth}
\centering
\includegraphics[width=\textwidth]{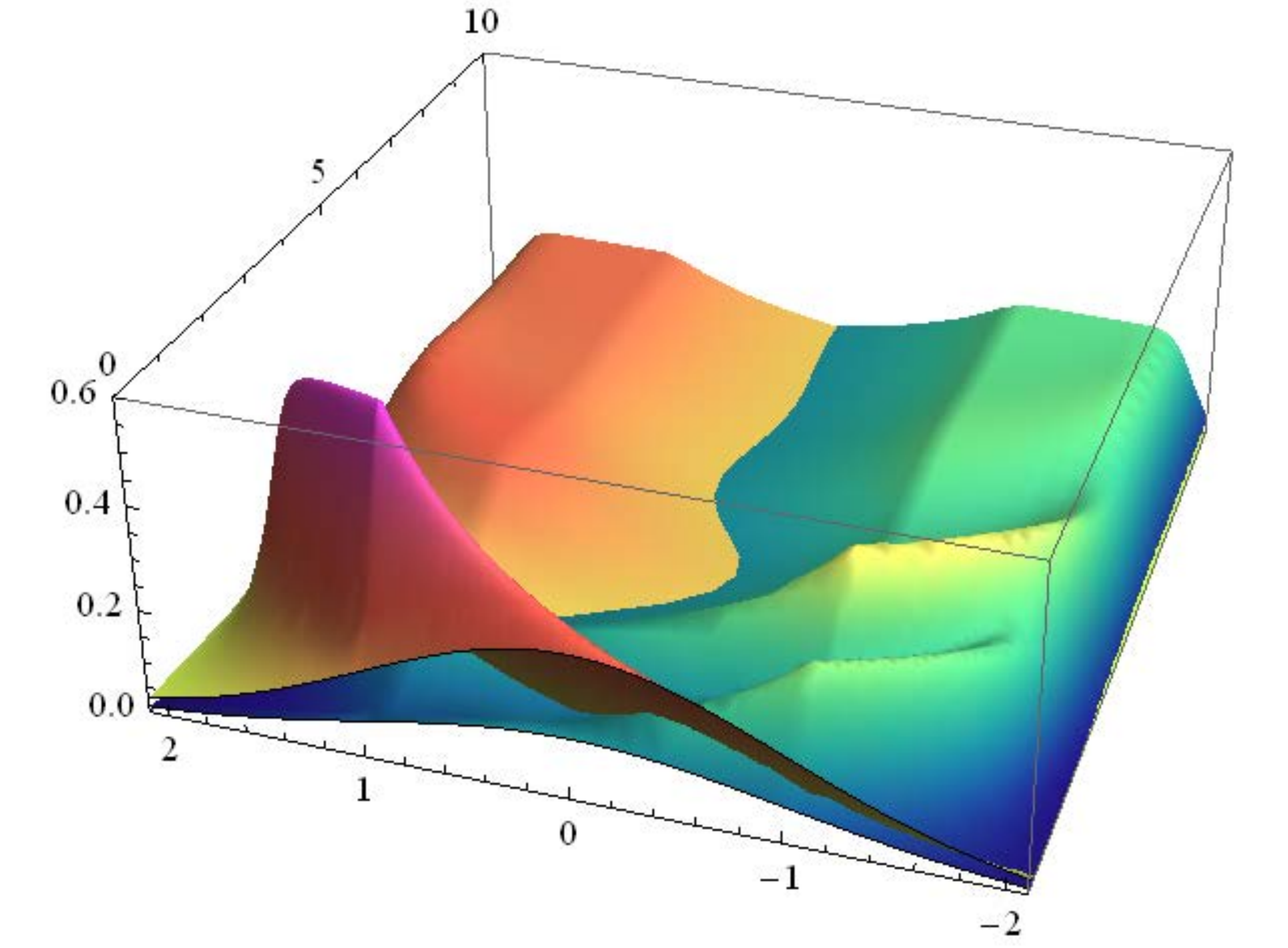}
\caption{$r=1$.}
\end{subfigure}
\begin{subfigure}{0.3\textwidth}
\centering
\includegraphics[width=\textwidth]{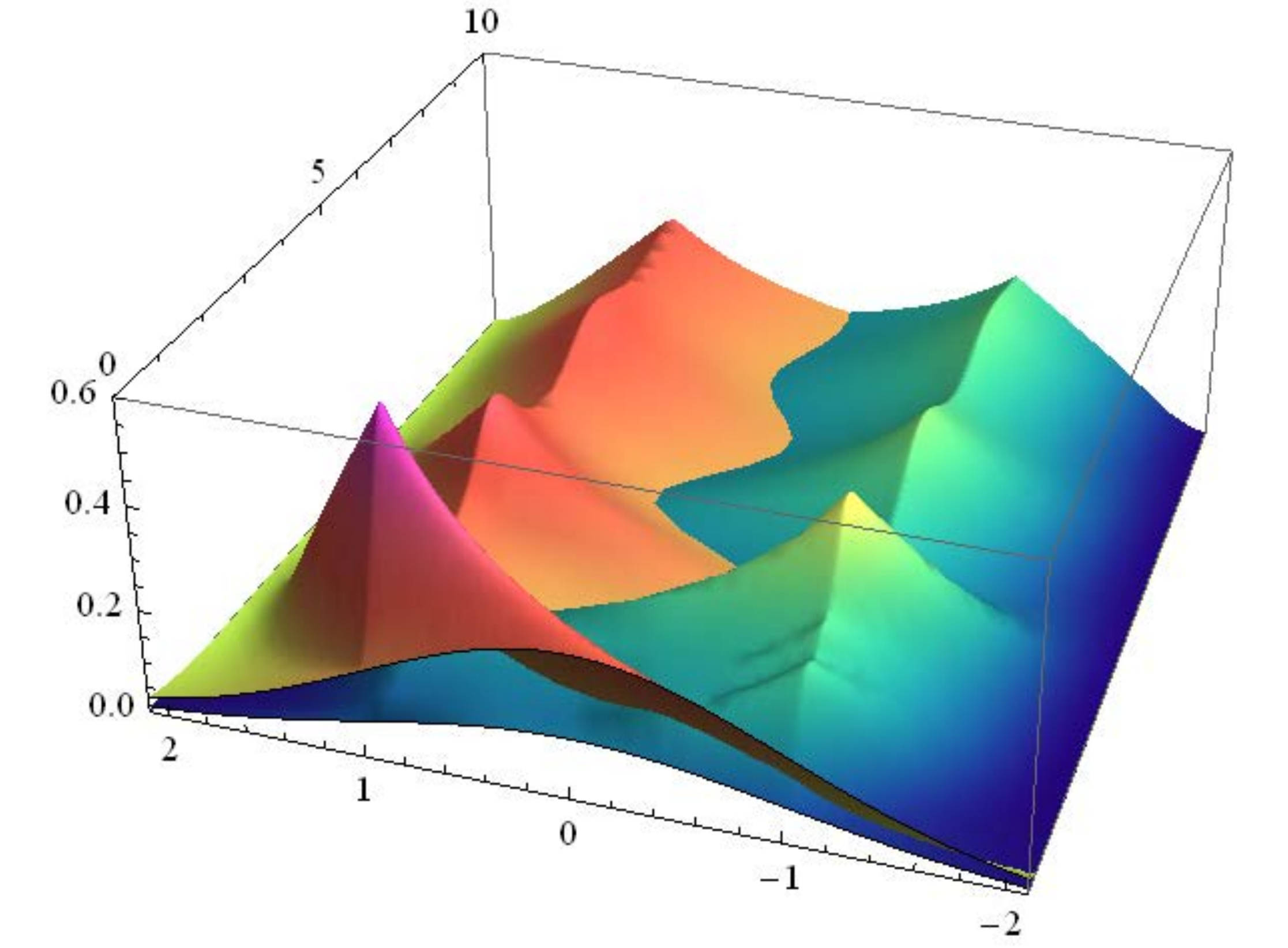}
\caption{$r=2$.}
\end{subfigure}
\begin{subfigure}{0.3\textwidth}
\centering
\includegraphics[width=\textwidth]{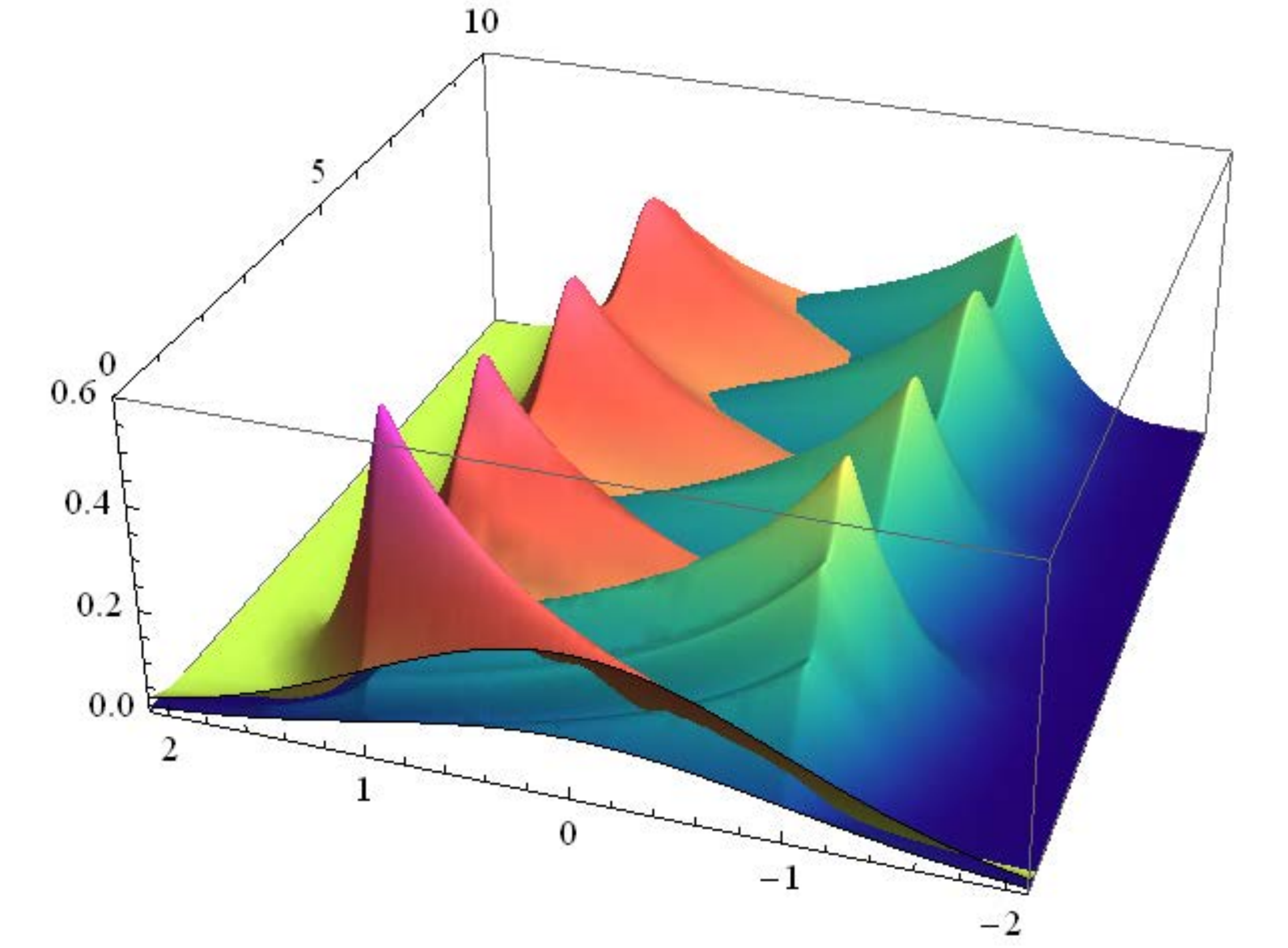}
\caption{$r=4$.}
\end{subfigure}
\caption{{\bf Evolution of the PDF for {\bf on}/{\bf off} (blue/red) states} shown as a function of temperature, $x\in[-2,2]$, and time, $t\in[0,10]$. The initial distributions are chosen Gaussian with $0.7/0.3$ proportion of devices between the {\bf on}/{\bf off} states. $\tau=1$, $\kappa=0.01$.}
\label{fig:on_off}
\end{figure}

To validate and extend the results of the theoretical analysis described in the two preceding Sections, we perform direct numerical simulation of the FP Eqs.~(\ref{FP-two-state_1}). The results are shown in Figures (\ref{fig:on_off}). We fix $\tau=1$, i.e. measure all other temporal characteristics in the units of $\tau$,  and choose for the experiments the $x$/temperature such that all relative temperatures are of the same order, specifically $x_\downarrow=-1, x_-=-1, x_+=1, x_\uparrow=1$. We also choose diffusion to be relatively small, $\kappa=0.01$, in accordance with what one expects to be of interest in practical applications. Four three dimensional Subfigures of Figure (\ref{fig:on_off}) show evolution of the PDFs in time starting from the initial PDFs chosen in the form of two Gaussians centered at $x=0$ and split in the $0.7/0.3$ proportion between the the {\bf on} and {\bf off} states
Main features of the PDFs dynamics seen in the simulations are as follows:
\begin{itemize}
\item When $r$ is sufficiently small the {\bf on} and {\bf off} ensembles do not mix initially equilibrating internally to the distributions picked a bit to the right/left from $x_-$/$x_+$ within $O(\tau)$ time.  Mixing between ensembles leading to establishment of a steady distribution is seen in $0(1/r)$ time. This final relaxation to the steady state is of a pure decay type.

\item We observe oscillations in the transients when $r$ is increased.  Some first signs of the oscillations are seen at $r<1$.

\item A transition in behavior of the PDF is observed at $r=1$. For example for the {\bf on} state, PDFs increases/decreases with $x$ decrease at $x_-<x< x_\downarrow$ at $r\tau<1$/$r\tau>1$. This is seen as a gradual shift of the {\bf on} state PDF center from $x_-$ to $x_\uparrow$ with $r$ increase.

\item Oscillations mature and become well pronounced at $r>1$. Oscillations stop to decay with further increase of $r$, turning asymptotically at $r\to\infty$ into a perfect oscillatory evolution with the period $t_{dc}$.
\end{itemize}
Notice also that all the features listed above as observed in the direct simulations are fully consistent with the theoretical and numerical results of the spectral analysis discussed in the preceding Subsections.

\section{Lagrangian Representation: Dynamics of a Device}
\label{sec:Lagr_dyn}

Consider an individual device cycling in the $(x,\sigma)$ space, like a particle in physics moving in a physical space, e.g. in a fluid flow. We call this view Lagrangian to contrast it with the Eulerian view we have followed so far in the manuscript representing instantaneous probability distribution over the whole $(x,\sigma)$ space. The new Lagrangian look at the dynamics brings new objects. In particular, we will analyze statistics of the number of cycles made by a device in the $(x,\sigma)$ space  in a finite time $t$, and show that this object is directly related the spectrum of the FP operator.  Such a straight relation between the two, normally unrelated (in statistical physics) objects, is unusual.

Material in this Section will be presented in three steps. First, in Section \ref{subsec:dyn_det} we analyze purely deterministic Lagrangian cycling of a device in the $(x,\sigma)$ space subject to the bang-bang, $r=+\infty$, control. Then, in Section \ref{subsec:dyn_jumps} we consider the case of a finite $r$ where the periodicity/determinism is broken and one will discuss statistics of cycling. Specifically, we introduce statistics of the flux and relate it to the discrete spectrum of the FP operator analysed in Section \ref{sub:soft_zero_dif}. Finally, in Section \ref{subsec:dyn_jumps+stoch},  we discuss how accounting for the Langevin/stochastic perturbations modify the picture.

\subsection{Deterministic Cycling: $r=+\infty$, $\kappa=0$.}
\label{subsec:dyn_det}

In this case the device motion is limited to the $[x_\downarrow,x_\uparrow]$ interval. If initially the device was at the {\bf on} state at $x_\uparrow$ its temperature decreases exponentially, according to
$\log\frac{x_\uparrow-x_-}{x(t)-x_-}=t/\tau$. At $t_+=\log\frac{x_\uparrow-x_-}{x_\downarrow-x_-}$, when the device temperature reaches, $x_\downarrow$, it switches to the {\bf off} state, entering the stage where temperature increases according to, $\log\frac{x_+-x_\uparrow}{x_+-x(t)}=(t-t_+)/\tau$. At $t_{dc}$, defined in Eq.~(\ref{t_dc}), the device reaches $x_\uparrow$, then transitions to the {\bf on} state, therefore completing the cycle.

The deterministic dynamics means, in particular, lack of mixing within the ensemble of the devices in the $r=\infty,\kappa=0$ case, when the dynamics simply carry the initial distribution of the ensemble returning it back to exactly the same initial distribution at every $t=n t_{dc}$, where $n$ is a positive integer.

When $\kappa$ is small but nonzero mixing will eventually take over.  In this case the resulting stationary PDFs in $x$ space correspond to uniform distribution of the devices in time, i.e. $P_{\uparrow/\downarrow}(x)\propto \tau/|x-x_\pm|$.

\subsection{Statistics of Cycles and Beyond. Finite $r$, $\kappa=0$.}
\label{subsec:dyn_jumps}

Stochasticity, and thus mixing, is brought in the difusionless model by Poisson switching between {\bf on}/{\bf off} levels. In this case two independent, Poisson distributed intervals, and also additional traveling times, when the device is outside of the $[x_\downarrow,x_\uparrow]$ interval/level, should be added to $t_{osc}$ to estimate the overall cycling time. Let the Poisson distributed time for the device to transition from {\bf on} to {\bf off} state, the overall time for the device to be outside of the comfort zone during the {\bf on} to {\bf off} transition, and the position
where the device switches from the {\bf on} state to the {\bf off} state, be $t$, $t_{\mbox{\scriptsize out}}$ and $x$ respectively. Then the three characteristics are related to each other according to
\begin{eqnarray}
&& t=\tau \log\left(\frac{x_\downarrow-x_-}{x-x_-}\right),\quad t_{\mbox{\scriptsize out}}-t=\tau \log\left(\frac{x_+-x}{x_+-x_\downarrow}\right).\label{x_t_t_out}
\end{eqnarray}
Excluding $x$ from Eqs.~(\ref{x_t_t_out}) one arrives at
\begin{eqnarray}
t_{\mbox{\scriptsize out}}=t+\tau \log\left(1+\alpha(1-e^{-t/\tau})\right), \quad \alpha\doteq \frac{x_\downarrow-x_-}{x_+-x_\downarrow}.
\label{t_out}
\end{eqnarray}
Inverting the relation (\ref{t_out}) one derives
\begin{eqnarray}
t=t_{\mbox{\scriptsize out}}+\tau\log \frac{1+\alpha e^{-t_{\mbox{\scriptsize out}}/\tau}}{1+\alpha}.
\label{t-via-t-out}
\end{eqnarray}
Then PDF of being out of the comfort zone for the time $t$ during the transition from {\bf on} to {\bf off} becomes
\begin{eqnarray}
P_{\mbox{\scriptsize out;down}}(t_{\mbox{\scriptsize out}})=r \exp(-r t) \frac{dt}{dt_{\mbox{\scriptsize out}}}=\frac{r}{1+\alpha e^{-t_{\mbox{\scriptsize out}}/\tau}} \left(\frac{1+\alpha}{e^{t_{\mbox{\scriptsize out}}/\tau}+\alpha}\right)^{r\tau}.
\label{P_out_down}
\end{eqnarray}
The analog of Eq.~(\ref{P_out_down}) for the transition from {\bf off} to {\bf on} is
\begin{eqnarray}
P_{\mbox{\scriptsize out;up}}(t_{\mbox{\scriptsize out}})=\frac{r}{1+\beta e^{-t_{\mbox{\scriptsize out}}/\tau}}\left(\frac{1+\beta}{e^{t_{\mbox{\scriptsize out}}/\tau}+\beta}\right)^{r\tau},\quad \beta\doteq \frac{x_+-x_\uparrow}{x_\uparrow-x_-}.
\label{P_out_up}
\end{eqnarray}
Finally, the PDF of being outside of the comfort zone for the time $t$ after completing $n$ cycles is
\begin{eqnarray}
&& P_{\mbox{\scriptsize{out};n}}(t)=\int\limits_{-i\infty+\epsilon}^{+i\infty +\epsilon}\frac{ds}{2\pi i} e^{s t} \left(F_s(\alpha) F_s(\beta)\right)^n,
\label{P_out_n}\\
&& F_s(\alpha)\doteq \int\limits_0^\infty dt e^{-t s}\frac{r}{1+\alpha e^{-t/\tau}}\left(\frac{1+\alpha}{e^{t/\tau}+\alpha}\right)^{r\tau}.
\label{F_s_alpha}
\end{eqnarray}

PDF for a device to spend time $t$ for making $n$ cycles is $P_{\mbox{\scriptsize{out};n}}(t-n t_{dc})$, where we just accounted for the fact that the total time of device's dynamics is summed up from the time to be outside of the comfort zone and the deterministic time $t_{dc}$ of moving through the cycle within the comfort zone.  Then PDF for the device to make $n$ cycles in time $t$ is recomputed according to the Bias formula
\begin{eqnarray}
P(n|t)=\frac{P_{\mbox{\scriptsize{out};n}}(t-n t_{dc})}{\sum_n  P_{\mbox{\scriptsize{out};n}}(t-n t_{dc})}.\label{Bias}
\end{eqnarray}
Interested to study statistics of the flux, $\omega$, defined as number of cycles per observation time $t$, $n=\omega t$,
we substitute Eq.~(\ref{P_out_n}) in Eq.~(\ref{Bias}). Evaluating the resulting integrals in the saddle point approximation and keeping only
the leading asymptotic terms one arrives at the following asymptotic Large Deviation (LD) estimate for the finite time PDF of the flux
\begin{eqnarray}
&& P(\omega|t) \sim \int\limits_{-i\infty+\epsilon}^{+i\infty +\epsilon}\frac{ds}{2\pi i} \exp\left(s t+\omega \log G_s
\right)\sim \exp\left(-t S(\omega)\right),
\label{P_omega_t}\\
&& G_s=(\alpha\beta)^{\tau s}F_s(\alpha) F_s(\beta),\label{G_s}\\
&& S(\omega)=-s_*-\omega G_{s_*},\quad 1=-\omega \left.\frac{d}{ds} \log(G_s)\right|_{s=s_*}.
\label{S_omega}
\end{eqnarray}
Therefore, we have an implicit expression for $S(\omega)$, the so-called LD function, via a newly introduced and explicitly known, $G_s$, function.
We will see shortly that $G_s$ is directly related to the spectrum of the FP operator.

\subsubsection{
Relating Eulerian and Lagrangian}
\label{subsub:EurLagr}

Explicit expressions derived above for the probability of advancing along the cycle in time $t$ can also be used to compute an Eulerian object -- the probability of observing  device in the state $(x,\sigma)$ at a time $t$. This is achieved relating the probability $P(x,\sigma|x_\uparrow,\uparrow;t)$ of the system to be in state $(x, \sigma)$ at time $t$, conditioned to be in state $(x_{0}, \sigma_{0})$ initially,
to the probability $\hat{P}(t|x_0, \sigma_{0} \to x,\sigma)$ of reaching the state $(x, \sigma)$, starting at $(x_{0}, \sigma_{0})$, in time $t$.
The former object is
\begin{eqnarray}
\label{Jacobian-t-to-x} P(x,\sigma|x_{0}, \sigma_{0};t) = \frac{1}{\dot{x}(x, \sigma)}\hat{P}(t|x_0, \sigma_{0} \to x,\sigma),
\end{eqnarray}
where $\dot{x}(x, \uparrow) = -(x - x_{-})/\tau$, and $\dot{x}(x, \downarrow) = -(x - x_{+})/\tau$, is the velocity of the deterministic motion at point $x$ on the discrete level $\sigma$; and the latter object is
\begin{eqnarray}
\hat{P}(t | x_\uparrow,\uparrow \to x, \sigma)=\sum_{n=0}^{m=\lfloor t/t_{dc}\rfloor} \int_0 ^{t-n t_{dc}}d t'  P(t'|x_\uparrow,\uparrow\to x,\sigma) P_{\mbox{\scriptsize{out};n}}(t-n t_{dc}-t'),
\label{P_x_sigma_t}
\end{eqnarray}
where $t_{dc}$, defined in Eq.~(\ref{t_dc}), is the time needed for the device to complete one deterministic cycle within the comfort zone. $P(t'|x_0, \sigma_{0} \to x,\sigma)$, entering Eq.~(\ref{P_x_sigma_t}) is the probability for the device to transition from the state $(x_{0}, \sigma_{0})$ to the state $(x,\sigma)$ in time $t'$ without completing a single cycle. It is assumed in Eq.~(\ref{P_x_sigma_t}) that all the devices are in the $(x_\uparrow,\uparrow)$ state initially and the choice of the initial state is made to simplify the resulting expressions. Notice, that at $t\to\infty$ the initial state will be forgotten, assuming sufficient mixing.

It is important to stress that Eq.~(\ref{Jacobian-t-to-x}) plays a crucial technical role in our approach by relating the distribution of $x$ in the l.h.s., which is the object of our interest, to the distribution of time in the r.h.s., which is easy to compute, using the Laplace transform technique. Such a simple relation via just a Jacobian is due to the deterministic nature of dynamics for a given discrete level, with the jumps between the discrete levels being the only source of stochasticity.

Obviously, Eq.~(\ref{P_x_sigma_t}) simplifies when $(x,\sigma)$ is chosen to coincide with the the initial state, $(x_\uparrow,\uparrow)$. We will focus on this case, where $P(t'|x_\uparrow,\uparrow\to x_\uparrow,\uparrow)=\delta(t')$ and thus Eq.~(\ref{P_x_sigma_t}) becomes
\begin{eqnarray}
P(x_\uparrow,\uparrow|x_\uparrow,\uparrow;t)= u^{-1}\sum_{n=0}^{m=\lfloor t/t_{dc}\rfloor} P_{\mbox{\scriptsize{out};n}}(t-n t_{dc}),
\label{P_x_sigma_t_simpl}
\end{eqnarray}
with $u = |\dot{x}(x_{\uparrow}, \uparrow)|$.

Extending $P_{\mbox{\scriptsize{out};n}}(t)$ with zero to negative times, $t < 0$, we can substitute $m$ in Eq.~(\ref{P_x_sigma_t_simpl}) by $\infty$.
Then, the Laplace transform (over time) version of Eq.~(\ref{P_x_sigma_t_simpl}) becomes
\begin{eqnarray}
P_s(x_\uparrow,\uparrow|x_\uparrow,\uparrow)&=& u^{-1}\int\limits_0^\infty dt e^{-t s} P(x_\uparrow,\uparrow|x_\uparrow,\uparrow;t) =
u^{-1}\sum_{n=0}^\infty \left(F_s(\alpha) F_s(\beta)\right)^n \exp(-s n t_{dc})\nonumber\\ && = \frac{u^{-1}}{1-F_s(\alpha) F_s(\beta)\exp(-s t_{dc})}=\frac{u^{-1}}{1-F_s(\alpha) F_s(\beta)(\alpha\beta)^{s\tau}},
\label{P_x_sigma_t_simpl_laplace}
\end{eqnarray}
given that $F_s(\alpha)F_s(\beta)< 1$, where $F_s(\alpha)$ is defined in Eq.~(\ref{F_s_alpha}). Inverse Laplace transform applied to the r.h.s. of Eq.~(\ref{P_x_sigma_t_simpl_laplace})
\begin{eqnarray}
P(x_\uparrow,\uparrow|x_\uparrow,\uparrow;t) = u^{-1}\int\limits_{-i\infty+\epsilon}^{+i\infty +\epsilon}\frac{ds}{2\pi i} e^{s t}
\frac{1}{1-G_s},
\label{P_x_sigma_t_simpl_2}
\end{eqnarray}
results in a sum over poles (in the domain of complex $s$) solving
\begin{eqnarray}
1=G_s=(\alpha\beta)^{s\tau} F_s(\alpha)F_s(\beta).
\label{poles}
\end{eqnarray}

Following the same strategy and with a bit of additional efforts an explicit expression for $P(x, \sigma | x_{0}, \sigma_{0}; t)$ for a full range of arguments can be derived. Indeed, introducing the notation
\begin{eqnarray}
\label{define-premodes} \xi_{s}(x, \sigma) = \frac{1}{|\dot{x}(x, \sigma)|}\int_{0}^{\infty}dte^{-st}P(t | x_{\uparrow}, \uparrow \to x, \sigma), \;\;\; \eta_{s}(x_{0}, \sigma_{0}) = \int_{0}^{\infty}dte^{-st}P(t | x_{0}, \sigma_{0} \to x_{\uparrow}, \uparrow),
\end{eqnarray}
we arrive at the following extended version of Eq.~(\ref{P_x_sigma_t_simpl_2})
\begin{eqnarray}
\label{P_x_sigma_t_simpl_2-extend} P(x, \sigma | x_{0}, \sigma_{0}; t) = \int_{-i\infty + \epsilon}^{i\infty + \epsilon}\frac{ds}{2\pi i}e^{st}\eta_{s}(x_{0}, \sigma_{0})\frac{1}{1-G_{s}}\xi_{s}(x, \sigma) + \tilde{P}(x, \sigma | x_{0}, \sigma_{0}; t),
\end{eqnarray}
with $\tilde{P}(x, \sigma | x_{0}, \sigma_{0}; t)$ being the PDF of being in state $(x, \sigma)$ at time $t$ provided the starting state was $(x_{0}, \sigma_{0})$ and a stochastic trajectory never went through state $(x_{\uparrow}, \uparrow)$. A direct computation yields the following natural relation for the eigenmodes $\xi_{\sigma, n}(x) = \xi_{-\lambda_{n}}(x, \sigma)$ and $\eta_{\sigma, n}(x) = \eta_{-\lambda_{n}}(x, \sigma)$.

Three follow up remarks are in order.

First, we note that Eq.~(\ref{poles}) is fully consistent with the spectral expression (\ref{zero-diff-spectral}) when $s\to -\lambda_n$. To check the consistency one just needs to change variables in the first and second integrals on the left hand side of Eq.~(\ref{zero-diff-spectral}) according to, $x=\frac{x_+\alpha +x_- \exp(t/\tau)}{\alpha+\exp(t/\tau)}$, and,
$x=\frac{x_-\beta +x_+ \exp(t/\tau)}{\beta+\exp(t/\tau)}$, respectively. In fact, the dynamical derivation of Eq.~(\ref{poles}) just presented should be considered as a rigorous proof of the spectral assumptions made  in Section \ref{subsub:zero_diff}.

Second, we observe that according to Eqs.~(\ref{G_s},\ref{S_omega},\ref{poles}) and preceding remark the LD function, $S(\omega)$, is
directly related to, $G_s$, fully defining the spectrum of the FP operator.

Finally, the sum over poles in our description is finite.  We do not prove this fact here but instead  illustrate it in Appendix \ref{sec:soft-zero-noise-spectral-toy} on a simpler toy model assuming an "instantaneous escape" (from the uncomfortable zone).

\subsection{General case: finite $r$, nonzero $\kappa$.}
\label{subsec:dyn_jumps+stoch}

We postpone discussion of the general case to future publication and will only make here some high level comments.

It is obvious that the dynamic analysis can also be extended to the stochastic case. In this case, we will need to account for stochastic/Langevin nature of the intrinsic, i.e. within the {\bf on} and {\bf off} states, dynamics.  Evaluating Poisson distributed transitions between the states we will also need to account for the fact that the Poisson time clock will need to be restarted each time the stochastic dynamics move the device from the comfort state (within the $[x_\downarrow;x_\uparrow]$ range) to the discomfort state. This is straightforward via the first passage technique of the stochastic calculus. (See also Appendix \ref{sec:relax-flux-hard} for a brief discussion on the application of the first passage approach to the case of the hard model.)

\section{Conclusions \& Path Forward}
\label{sec:conclusions}

This paper presents detailed analysis of the model accounting for stochastic dynamics of typical thermostatic devices cycling in the mixed state describing temperature and the switch (on or off) status of the device. Switching is modeled as a random Poisson process.  We consider ensemble of similar devices and study dynamics of the probability distribution function of the mixed state (temperature and {\bf on}/{\bf off} status) in time. FP equations for temporal evolution of the PDF in space and time were derived and analyzed by means of the spectral (Eulerian) and dynamic (Lagrangian) analysis. In particular,  we show that the spectral analysis is reduced to solving a trancendental equation on the eigenvalue. The equation is analyzed analytically in limiting cases and otherwise it is studied numerically.  A particularly interesting consequence of this analysis is establishment of the fact that Poisson transitions are sufficient for mixing the system efficiently even in the case of zero thermal diffusivity. Our analysis yields detailed results and intuitive explanations for how the system evolve in time approaching the steady state. We also analyze PDF of the finite time flux, defined as the number of the phase space cycles made by a device in a fixed time. We show that the long time asymptotic of this object can be reconstructed directly from the spectrum of the Fokker-Planck operator. This relation is akin to relation between Eulerian (instantaneous velocity) and Lagrangian (particle dynamics) description in physics, e.g. in fluid mechanics.

Results of the paper are of importance to both engineering (including theoretical engineering, power system engineering and more generally energy systems engineering) and statistical physics communities.

Main consequence of this paper analysis on the field of engineering is in establishing dependencies of the relaxation/mixing rate of the ensemble on the parameters of the model, especially on switching {\bf on}/{\bf off} rate proposed as the major characteristic providing control of the TCL ensembles \cite{11CH,13MKC,15MKLAC}.
Of a particular interest generalization of our results to control of more realistic ensembles,
e.g. to heterogeneous/inhomogeneous ensembles including devices with different characteristics of the type discussed in \cite{15GMK}, or even more generally coarse-grained ensembles described within the Markov Decision Process (MDP) framework binned/discretized in space (temperature or other exogenous characteristics) and, possibly, time, see e.g. \cite{15MBBCE,15PKL}. Note that we discuss a related MDP approach, e.g. taking advantage of the linearly-solvable control problems \cite{82FM,12DjEmo,15MBBCE}, in a companion paper \cite{17CCb}.

Models and methods of their analysis reported in the paper are also important for the field of statistical physics due to an unusual combination of, on one hand, description of a quintessential non-equilibrium problems with detailed balance strongly violated, while on the other hand being analytically solvable.  Indeed even in the equilibrium statistical mechanics,  where steady solution of the FP equation has a close form Gibbs form, spectrum of the FP operator, describing PDF dynamics, in general do not allow an explicit solution which is known only for a rather limited class of problems. Ability to find a steady state solution extends to some special class of non-equilibrium problems such as queuing networks of operations research \cite{63Jac,76Kel} and zero-range models \cite{70Spi,93DEM} of statistical physics, where one can also analyze statistics of the measure of detailed balance violation (which can be expressed as current, entropy or work produced) \cite{10CCGT}. However, in these known solvable non-equilibrium statistical physics problems, as in a general equilibrium mechanics problem, finding the entire spectrum of the FP operator,  or even its eigenvalue with the lowest nonzero real part, doomed impossible. Remarkably the soft model introduced and analyzed in this paper, in addition to being non-equilibrium problem where steady solution and statistics of current (measuring degree of the detailed balance violation) are known, also allow explicit closed form expression for the spectrum. Moreover, we show that spectrum of the FP operator in this special mixed problems is related explicitly to statics of the current  Extending this ``complete" non-equilibrium solvability to other areas of theoretical and applied statistical mechanics, such as statistical hydrodynamics and thus complementing body of work on analytically tractable models of turbulence, akin passive scalar and burgulence theories, see e.g. \cite{01FGV} for a review, would thus be of a great interest.

\begin{acknowledgments}
The authors are grateful to S. Backhaus, I. Hiskens, D. Calloway for fruitful discussions and valuable comments. The work at LANL was carried out under the auspices of the National Nuclear Security Administration of the U.S. Department of Energy under Contract No. DE-AC52-06NA25396.
\end{acknowledgments}

\appendix

\section{Details of Spectral Computations for Hard Model}
\label{app:Hard-Spectral}

Substituting Eqs.~(\ref{eigen-exp}) into Eqs.~(\ref{FP-hard_1},\ref{FP-hard_2}) one finds that
$\xi_{\uparrow/\downarrow,n}$ satisfy
\begin{eqnarray}
\label{eigenvalue-left}  \kappa\partial_{x}^{2}\xi_{\uparrow,n}+ \partial_{x}\left(\frac{x-x_{-}}{\tau}\xi_{\uparrow,n}\right)-  \kappa(\partial_{x}\xi_{\downarrow,n}(x_{\uparrow}))\delta(x- x_{\uparrow}) &=& -\lambda_{n}\xi_{\uparrow,n}, \nonumber \\ \kappa\partial_{x}^{2}\xi_{\downarrow,n}+ \partial_{x}\left(\frac{x-x_{+}}{\tau}\xi_{\downarrow,n}\right)  + \kappa(\partial_{x}\xi_{\uparrow,n}(x_{\downarrow}))\delta(x- x_{\downarrow}) &=& -\lambda_{n}\xi_{\downarrow,n}.
\end{eqnarray}
Strictly within the "hard" range, $x\in ]x_\downarrow,x_\uparrow[$, general solution of the Eqs.~(\ref{eigenvalue-left}) becomes
\begin{eqnarray}
&& \xi_{\uparrow/\downarrow,n}(x)=c_{\uparrow/\downarrow,1,n} \xi_{\uparrow/\downarrow,1,n}(x)+c_{\uparrow/\downarrow,2,n}\xi_{\uparrow/\downarrow,2,n}(x) \label{hard}\\
&& \xi_{\uparrow/\downarrow,1,n}(x)\doteq
\exp\left(-\frac{x^2}{2\kappa\tau}+\frac{x x_{-/+}}{\kappa\tau}\right)
\, _1F_1\left(\!-\frac{\lambda_n\tau}{2},\frac{1}{2},\frac{(x-x_{-/+})^2}{2\kappa\tau}\!\right)
\label{hard_j1n}\\
&& \xi_{\uparrow/\downarrow,2,n}(x)\doteq
\exp\left(-\frac{x^2}{2\kappa\tau}+\frac{x x_{-/+}}{\kappa\tau}\right) (x-x_{-/+})
\, _1F_1\!\left(\!-\frac{\lambda_n\tau}{2}+\frac{1}{2},\frac{3}{2},\frac{(x-x_{-/+})^2}{2\kappa\tau}\!\right)
\label{hard_j2n}
\end{eqnarray}
where $c$ are yet to be defined constants and $\, _1F_1(a,b,x)$ is the Kummer's confluent hypergeometric function. It is important to emphasize that here in Eqs.~(\ref{hard}) we pick a pair of the homogeneous solutions which are linearly independent at all values of $\lambda$ and which are both not singular within the $[x_\downarrow,x_\uparrow]$ range. (Notice that even thought the second terms in Eq.~(\ref{hard}) are singular at $x=x_{-/+}$ the values are both outside of the domain of interest.) Conditions at the boundaries of the "hard" range (with the values taken at the inner sides of the range) follow from direct integrations of Eqs.~(\ref{hard}) over infinitesimally small domains around the boundary values supplemented with the extra condition that values of all the $\xi$ functions outside of the hard domain are zero. One arrives at
\begin{eqnarray}
\label{xi0} && \xi_{\uparrow,n}(x_\downarrow)=\xi_{\downarrow,n}(x_\uparrow)=0\\
\label{xin0} && \kappa \partial_x\xi_{\uparrow/\downarrow,n}(x_{\uparrow/\downarrow})+\frac{x_{\uparrow/\downarrow}-
x_{-/+}}{\tau}\xi_{\uparrow/\downarrow,n}(x_{\uparrow/\downarrow})+\kappa \partial_x\xi_{\downarrow/\uparrow,n}(x_{\uparrow/\downarrow})=0.
\end{eqnarray}
Substituting Eqs.~(\ref{hard}) into Eqs.~(\ref{xi0},\ref{xin0}) we arrive at four linear homogeneous relations on four $c$-constants:
\begin{eqnarray}
&& M(\lambda_n)\left(\begin{array}{c} c_{\uparrow,1,n}\\ c_{\uparrow,2,n}\\ c_{\downarrow,1,n}\\ c_{\downarrow,2,n}\end{array}\right)=0,
\label{M-c}\\
&& \hspace{-0.1cm} M\!\doteq\!\! \left(\!
\scalemath{0.7}{
\begin{array}{cccc}
\xi_{\uparrow,1,n}(x_\downarrow) & \xi_{\uparrow,2,n}(x_\downarrow) & 0 & 0\\
0 & 0 & \xi_{\downarrow,1,n}(x_{\downarrow}) & \xi_{\downarrow,2,n}(x_{\downarrow})\\
\kappa \partial_x{\xi}_{\uparrow,1,n}(x_{\downarrow})+\frac{x_{\downarrow}-x_-}{\tau}\xi_{\uparrow,1,n}(x_{\downarrow}) &
\kappa \partial_x{\xi}_{\uparrow,2,n}(x_{\downarrow})+\frac{x_{\downarrow}-x_-}{\tau}\xi_{\uparrow,2,n}(x_{\downarrow}) &
\kappa \partial_x{\xi}_{\downarrow,1,n}(x_{\downarrow}) & \kappa \partial_x{\xi}_{\downarrow,2,n}(x_{\downarrow})\\
\kappa \partial_x{\xi}_{\uparrow,1,n}(x_\downarrow) & \kappa \partial_x{\xi}_{\uparrow,2,n}(x_\downarrow) &
\kappa \partial_x{\xi}_{\downarrow,1,n}(x_\downarrow)+\frac{x_\downarrow-x_+}{\tau}\xi_{\downarrow,1,n}(x_\downarrow) &
\kappa \partial_x{\xi}_{\downarrow,2,n}(x_\downarrow)+\frac{x_\downarrow-x_+}{\tau}\xi_{\downarrow,2,n}(x_\downarrow)
\end{array}
}
\!\right)
\label{M_hard_explicit}
\end{eqnarray}
where $M(\lambda)$ is thus an explicitly defined $4\times 4$ matrix with coefficients parametrically dependent on $\lambda$. The allowed  $\lambda_n$ are solutions of the zero-determinant condition for the respective matrix,
\begin{eqnarray}
\mbox{det} M(\lambda)=0. \label{detM}
\end{eqnarray}
Even though the determinant is an explicit function of $\lambda$, resolving the zero determinant equation explicitly does not seems feasible and we will rely here on a numerical exploration of the zero-determinant Eq.~(\ref{detM}), which is illustrated in Fig.~(\ref{fig:hard-det}).

Notice also (for the sake of completeness), that the eigenvalue problem was analyzed in \cite{09Cal}, where the author guessed that $\lambda_n=2n/\tau$ making the first argument in the Kumar's confluent  hypergeometric function a negative integer (in which case the hypergeometric functions are reduced to the polynomials). However, a direct check shows that Eq.~(\ref{detM}) is not satisfied for the guess at a finite $\kappa$.

\section{Spectral Properties of a Toy "Instantaneous Escape" Model}
\label{sec:soft-zero-noise-spectral-toy}

\begin{figure}[t]
\centering
\includegraphics[width=0.3\textwidth]{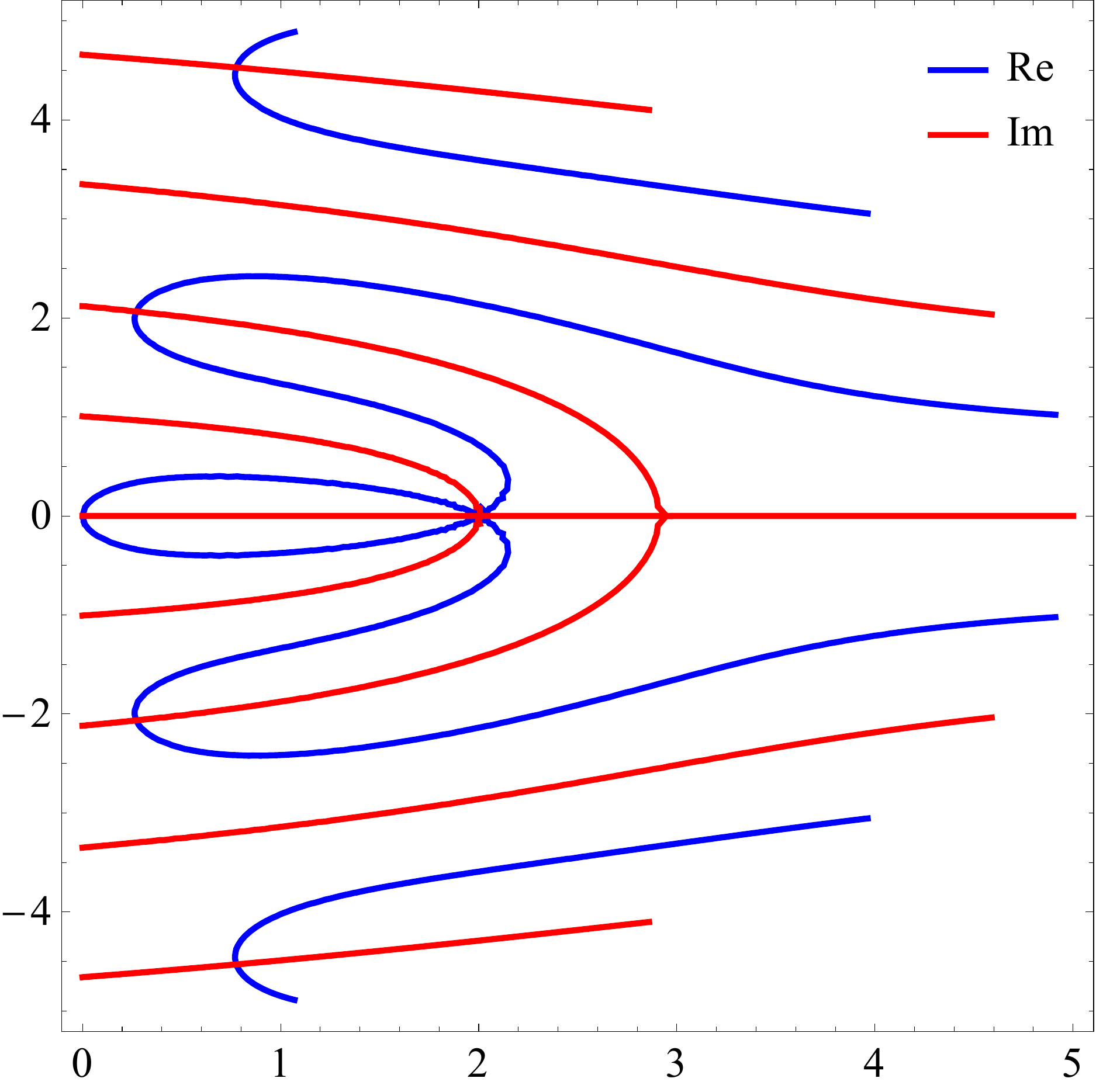}
\caption{Graphical solution of the spectral problem for the toy  "instantaneous escape" (from the uncomfortable zone) model discussed in Appendix \ref{sec:soft-zero-noise-spectral-toy}. We choose $\alpha$ and $\beta$ parameters as described before. Besides, we choose $\tau=1,\ r=2$.
\label{fig:simple_model}}
\end{figure}

In this Appendix we discuss spectrum in the simple case of a toy "instantaneous escape" (from the uncomfortable zone) model assuming that the device returns to the control zone instantaneously. In this case we have
\begin{eqnarray}
\label{G-of-lambda-toy} F_{s}(\alpha) = F_{s}(\beta) = F_{s} = -\frac{r}{r + s},
\end{eqnarray}
and the spectral Eq.~(\ref{poles}) simplifies to
\begin{eqnarray}
\label{spectrum-toy-2} (s + r)^{2} = r^{2}e^{-s t_{{\rm dc}}}.
\end{eqnarray}
One observes that at $r>r_{{\rm cr}}$ Eq.~(\ref{spectrum-toy-2}) has no real solutions, while at $r\le r_{\rm cr}$ there is exactly one real solution, where $r_{\rm cr}$ is given implicitly by
\begin{eqnarray}
\label{spectrum-toy-4} (r_{{\rm cr}}t_{{\rm dc}})^{2} = 4 e^{-(r_{{\rm cr}}t_{{\rm dc}} +2)}.
\end{eqnarray}
This implies, in particular, that $r_{{\rm cr}} < t_{{\rm dc}}^{-1}$.

Next let us analyze convergence of the spectral decomposition of the PDF of finding device in the stae $(x,\sigma)$ at time $t$, given that the device was in the state $(x_0,\downarrow)$ at the moment $0$
\begin{eqnarray}
\label{spectral-decomp} P(x, \downarrow | x_{0}, \downarrow; t) = \sum_{n}e^{-\varepsilon_{n} t-i\omega_n}\xi_{\downarrow, n}(x)\eta_{\downarrow, n}(x_{0}),
\end{eqnarray}
where $\xi$ and $\eta$ are respective left and right eigenvalues of the FP operator of the toy model.
Substituting $s$ by$-\lambda_n$ in Eq.~(\ref{spectrum-toy-2}) and analyzing the large $n$ part of the spectrum one derives
\begin{eqnarray}
\label{spectral-decomp-2} e^{-\lambda_{n} t} &\sim& e^{\pm i\pi (2n+1) (t/t_{{\rm dc}})}\left(\frac{rt_{{\rm dc}}}{\pi(2n+1)}\right)^{2(t/t_{{\rm dc}})}, \;\;\; \xi_{\downarrow ,n}(x) \sim e^{\pm i\pi (\tau/t_{{\rm dc}})(2n+1)\alpha_{\downarrow}(x)} \left(\frac{(2n+1)\pi}{rt_{{\rm dc}}}\right)^{2\alpha_{\downarrow}(x)\tau/t_{{\rm dc}}}, \nonumber \\ \eta_{\downarrow ,n}(x_{0}) &\sim& t_{{\rm dc}}^{-1} e^{\mp i\pi (\tau/t_{{\rm dc}})(2n+1)\alpha_{\downarrow}(x_{0})} \left(\frac{(2n+1)\pi}{rt_{{\rm dc}}}\right)^{-2\alpha_{\downarrow}(x_{0})\tau/t_{{\rm dc}}}, \;\;\; \alpha_{\downarrow}(x) = \ln \left(\frac{|x_{+} - x_{\downarrow}|}{|x_{+} - x|}\right).
\end{eqnarray}
It follows from  Eq.~(\ref{spectral-decomp-2}) that for given values of $x$ and $x_{0}$ the spectral decomposition of $P(x, \downarrow | x_{0}, \downarrow; t)$ converges at sufficiently large $t$. However, if the initial distribution $P_{\downarrow, 0}(x)$ is represented by a smooth function with compact support, the spectral decomposition converges at all times.

\section{Flux Generation in the Hard Model: Dynamical Viewpoint and First-Passage Time}
\label{sec:relax-flux-hard}

Even though relaxation dynamics of the hard ($r=\infty$) model and difusionless model are quite different the LD function of the hard model and its relation to the spectrum of the FP operator can be identified in the way similar to how it was done for the zero diffusivity model in Section~\ref{subsec:dyn_jumps}. An appropriate dynamic object in the case of the hard model is PDF of staying at the level $\sigma$ for time $t$, $P_{\sigma}(t)$, which we can also refer to as the PDF of the first-passage time at the level $\sigma$. PDF of making one full cycle in time $t$ becomes,
$P(t)=\int_0^t dt' P_\uparrow(t') P_\downarrow (t-t')$. Respective Laplace transforms
\begin{eqnarray}
\label{G-first-passage} G_{\sigma, s} = \int_{-\infty}^{\infty}dt e^{-st}P_{\sigma}(t),\quad G_{s} = \int_{-\infty}^{\infty}dt e^{-st}P(t)
\end{eqnarray}
are related to each other according to
\begin{eqnarray}
\label{G-cycle-time-gen} G_{s} = G_{\downarrow, s}G_{\uparrow, s},
\end{eqnarray}
which is the hard model version of Eq.~(\ref{G_s}). The respective LD function, $S(\omega)$, is obtained from Eqs.~(\ref{S_omega}), with $G_{s}$, taken the form of Eq.~(\ref{G-cycle-time-gen}).

Denote by $K_{\sigma}(x, x_{0}; t)$ solution of the FP Eqs.~(\ref{FP-hard_1},\ref{FP-hard_2}) with the absorbing boundary conditions (\ref{FP-hard-BC}), where thus the source term on the r.h.s. of the FP equations is substituted by the initial conditions $K_{\sigma}(x, x_{0}; 0) = \delta(x - x_{0})$. We then have
\begin{eqnarray}
\label{P-first-passage-expl} P_{\downarrow}(t) = -\kappa(\partial_{x}K_{\downarrow}(x, x_{\downarrow}; t))_{x=x_{\uparrow}}, \;\;\; P_{\uparrow}(t) = \kappa(\partial_{x}K_{\uparrow}(x, x_{\uparrow}; t))_{x=x_{\uparrow}}.
\end{eqnarray}
In the regime of weak diffusion, $\kappa \ll (x_{\uparrow} - x_{\downarrow})^{2}/\tau$, the first passage time distributions can be computed explicitly by using a spectral decomposition and utilizing WKB approximation of Quantum Mechanics to resolve the spectrum problem.

\bibliographystyle{naturemag}
\bibliography{TCL}

\begin{thebibliography}{10}
\expandafter\ifx\csname url\endcsname\relax
  \def\url#1{\texttt{#1}}\fi
\expandafter\ifx\csname urlprefix\endcsname\relax\def\urlprefix{URL }\fi
\providecommand{\bibinfo}[2]{#2}
\providecommand{\eprint}[2][]{\url{#2}}

\bibitem{79CD}
\bibinfo{author}{Chong, C.~Y.} \& \bibinfo{author}{Debs, A.~S.}
\newblock \bibinfo{title}{Statistical synthesis of power system functional load
  models}.
\newblock In \emph{\bibinfo{booktitle}{Decision and Control including the
  Symposium on Adaptive Processes, 1979 18th IEEE Conference on}},
  vol.~\bibinfo{volume}{2}, \bibinfo{pages}{264--269} (\bibinfo{year}{1979}).

\bibitem{81IS}
\bibinfo{author}{Ihara, S.} \& \bibinfo{author}{Schweppe, F.}
\newblock \bibinfo{title}{Physically based modeling of cold load pickup}.
\newblock \emph{\bibinfo{journal}{Power Apparatus and Systems, IEEE
  Transactions on}} \textbf{\bibinfo{volume}{PAS-100}},
  \bibinfo{pages}{4142--4150} (\bibinfo{year}{1981}).

\bibitem{84CM}
\bibinfo{author}{Chong, C.-Y.} \& \bibinfo{author}{Malhami, R.~P.}
\newblock \bibinfo{title}{Statistical synthesis of physically based load models
  with applications to cold load pickup}.
\newblock \emph{\bibinfo{journal}{Power Apparatus and Systems, IEEE
  Transactions on}} \textbf{\bibinfo{volume}{PAS-103}},
  \bibinfo{pages}{1621--1628} (\bibinfo{year}{1984}).

\bibitem{85MC}
\bibinfo{author}{Malhame, R.} \& \bibinfo{author}{Chong, C.-Y.}
\newblock \bibinfo{title}{Electric load model synthesis by diffusion
  approximation of a high-order hybrid-state stochastic system}.
\newblock \emph{\bibinfo{journal}{IEEE Transactions on Automatic Control}}
  \textbf{\bibinfo{volume}{30}}, \bibinfo{pages}{854--860}
  (\bibinfo{year}{1985}).

\bibitem{88MC}
\bibinfo{author}{Malhame, R.} \& \bibinfo{author}{Chong, C.-Y.}
\newblock \bibinfo{title}{On the statistical properties of a cyclic diffusion
  process arising in the modeling of thermostat-controlled electric power
  system loads}.
\newblock \emph{\bibinfo{journal}{SIAM Journal on Applied Mathematics}}
  \textbf{\bibinfo{volume}{48}}, \bibinfo{pages}{465--480}
  (\bibinfo{year}{1988}).
\newblock \urlprefix\url{http://dx.doi.org/10.1137/0148026}.
\newblock \eprint{http://dx.doi.org/10.1137/0148026}.

\bibitem{04LC}
\bibinfo{author}{Lu, N.} \& \bibinfo{author}{Chassin, D.}
\newblock \bibinfo{title}{A state-queueing model of thermostatically controlled
  appliances}.
\newblock \emph{\bibinfo{journal}{Power Systems, IEEE Transactions on}}
  \textbf{\bibinfo{volume}{19}}, \bibinfo{pages}{1666--1673}
  (\bibinfo{year}{2004}).

\bibitem{05LCW}
\bibinfo{author}{Lu, N.}, \bibinfo{author}{Chassin, D.} \&
  \bibinfo{author}{Widergren, S.}
\newblock \bibinfo{title}{Modeling uncertainties in aggregated thermostatically
  controlled loads using a state queueing model}.
\newblock \emph{\bibinfo{journal}{Power Systems, IEEE Transactions on}}
  \textbf{\bibinfo{volume}{20}}, \bibinfo{pages}{725--733}
  (\bibinfo{year}{2005}).

\bibitem{09Cal}
\bibinfo{author}{Callaway, D.~S.}
\newblock \bibinfo{title}{Tapping the energy storage potential in electric
  loads to deliver load following and regulation, with application to wind
  energy}.
\newblock \emph{\bibinfo{journal}{Energy Conversion and Management}}
  \textbf{\bibinfo{volume}{50}}, \bibinfo{pages}{1389 -- 1400}
  (\bibinfo{year}{2009}).
\newblock
  \urlprefix\url{http://www.sciencedirect.com/science/article/pii/S0196890408004780}.

\bibitem{11CH}
\bibinfo{author}{Callaway, D.} \& \bibinfo{author}{Hiskens, I.}
\newblock \bibinfo{title}{Achieving controllability of electric loads}.
\newblock \emph{\bibinfo{journal}{Proceedings of the IEEE}}
  \textbf{\bibinfo{volume}{99}}, \bibinfo{pages}{184--199}
  (\bibinfo{year}{2011}).

\bibitem{11BF}
\bibinfo{author}{Bashash, S.} \& \bibinfo{author}{Fathy, H.~K.}
\newblock \bibinfo{title}{Modeling and control insights into demand-side energy
  management through setpoint control of thermostatic loads}.
\newblock In \emph{\bibinfo{booktitle}{Proceedings of the 2011 American Control
  Conference}}, \bibinfo{pages}{4546--4553} (\bibinfo{year}{2011}).

\bibitem{12AK}
\bibinfo{author}{Angeli, D.} \& \bibinfo{author}{Kountouriotis, P.~A.}
\newblock \bibinfo{title}{A stochastic approach to dynamic-demand refrigerator
  control}.
\newblock \emph{\bibinfo{journal}{IEEE Transactions on Control Systems
  Technology}} \textbf{\bibinfo{volume}{20}}, \bibinfo{pages}{581--592}
  (\bibinfo{year}{2012}).

\bibitem{Feynman}
\bibinfo{author}{Feynman, R.~P.}
\newblock \emph{\bibinfo{title}{Statistical Mechanics}}
  (\bibinfo{publisher}{Advanced Books Classics, Perseus Books, Reading,
  Massachusets}, \bibinfo{year}{1997}).

\bibitem{vanKampen}
\bibinfo{author}{van Kampen, N.}
\newblock \emph{\bibinfo{title}{Stochastic Processes in Physics and Chemistry
  (Third Edition)}} (\bibinfo{publisher}{Amsterdam: Elsevier},
  \bibinfo{year}{2007}).

\bibitem{Gardiner}
\bibinfo{author}{Gardiner, C.~W.}
\newblock \emph{\bibinfo{title}{Handbook of stochastic methods for physics,
  chemistry and the natural sciences, 3rd ed.}} (\bibinfo{publisher}{Springer
  Series in Synergetics, vol.13, Berlin: Springer-Verlag},
  \bibinfo{year}{2004}).

\bibitem{15GMK}
\bibinfo{author}{Ghaffari, A.}, \bibinfo{author}{Moura, S.} \&
  \bibinfo{author}{Krstic, M.}
\newblock \bibinfo{title}{Pde-based modeling, control, and stability analysis
  of heterogeneous thermostatically controlled load populations}.
\newblock \emph{\bibinfo{journal}{Journal of Dynamic Systems Measurement and
  Control}} \textbf{\bibinfo{volume}{137}} (\bibinfo{year}{2015}).

\bibitem{13MKC}
\bibinfo{author}{Mathieu, J.}, \bibinfo{author}{Koch, S.} \&
  \bibinfo{author}{Callaway, D.}
\newblock \bibinfo{title}{State estimation and control of electric loads to
  manage real-time energy imbalance}.
\newblock \emph{\bibinfo{journal}{Power Systems, IEEE Transactions on}}
  \textbf{\bibinfo{volume}{28}}, \bibinfo{pages}{430--440}
  (\bibinfo{year}{2013}).

\bibitem{15MKLAC}
\bibinfo{author}{Mathieu, J.~L.}, \bibinfo{author}{Kamgarpour, M.},
  \bibinfo{author}{Lygeros, J.}, \bibinfo{author}{Andersson, G.} \&
  \bibinfo{author}{Callaway, D.~S.}
\newblock \bibinfo{title}{Arbitraging intraday wholesale energy market prices
  with aggregations of thermostatic loads}.
\newblock \emph{\bibinfo{journal}{IEEE Transactions on Power Systems}}
  \textbf{\bibinfo{volume}{30}}, \bibinfo{pages}{763--772}
  (\bibinfo{year}{2015}).

\bibitem{15MBBCE}
\bibinfo{author}{Meyn, S.~P.}, \bibinfo{author}{Barooah, P.},
  \bibinfo{author}{Busic, A.}, \bibinfo{author}{Chen, Y.} \&
  \bibinfo{author}{Ehren, J.}
\newblock \bibinfo{title}{Ancillary service to the grid using intelligent
  deferrable loads}.
\newblock \emph{\bibinfo{journal}{IEEE Transactions on Automatic Control}}
  \textbf{\bibinfo{volume}{60}}, \bibinfo{pages}{2847--2862}
  (\bibinfo{year}{2015}).

\bibitem{15PKL}
\bibinfo{author}{Paccagnan, D.}, \bibinfo{author}{Kamgarpour, M.} \&
  \bibinfo{author}{Lygeros, J.}
\newblock \bibinfo{title}{On the range of feasible power trajectories for a
  population of thermostatically controlled loads}.
\newblock In \emph{\bibinfo{booktitle}{2015 54th IEEE Conference on Decision
  and Control (CDC)}}, \bibinfo{pages}{5883--5888} (\bibinfo{year}{2015}).

\bibitem{82FM}
\bibinfo{author}{Fleming, W.~H.} \& \bibinfo{author}{Mitter, S.~K.}
\newblock \bibinfo{title}{Optimal control and nonlinear filtering for
  nondegenerate diffusion processes}.
\newblock \emph{\bibinfo{journal}{Stochastics}} \textbf{\bibinfo{volume}{8}},
  \bibinfo{pages}{63--77} (\bibinfo{year}{1982}).
\newblock \urlprefix\url{http://dx.doi.org/10.1080/17442508208833228}.
\newblock \eprint{http://dx.doi.org/10.1080/17442508208833228}.

\bibitem{12DjEmo}
\bibinfo{author}{{Dvijotham}, K.} \& \bibinfo{author}{{Todorov}, E.}
\newblock \bibinfo{title}{{A Unifying Framework for Linearly Solvable
  Control}}.
\newblock \emph{\bibinfo{journal}{ArXiv e-prints}}  (\bibinfo{year}{2012}).
\newblock \eprint{1202.3715}.

\bibitem{17CCb}
\bibinfo{author}{Chertkov, M.} \& \bibinfo{author}{Chernyak, V.}
\newblock \bibinfo{title}{Ensemble control of cycling energy loads: Markov
  decision approach} (\bibinfo{year}{2017}).

\bibitem{63Jac}
\bibinfo{author}{Jackson, J.~R.}
\newblock \bibinfo{title}{Jobshop-like queueing systems}.
\newblock \emph{\bibinfo{journal}{Management Science}}
  \textbf{\bibinfo{volume}{10}}, \bibinfo{pages}{131--142}
  (\bibinfo{year}{1963}).
\newblock \urlprefix\url{http://www.jstor.org/stable/2627213}.

\bibitem{76Kel}
\bibinfo{author}{Kelly, F.~P.}
\newblock \bibinfo{title}{Networks of queues}.
\newblock \emph{\bibinfo{journal}{Advances in Applied Probability}}
  \textbf{\bibinfo{volume}{8}}, \bibinfo{pages}{416--432}
  (\bibinfo{year}{1976}).
\newblock \urlprefix\url{http://www.jstor.org/stable/1425912}.

\bibitem{70Spi}
\bibinfo{author}{Spitzer, F.}
\newblock \bibinfo{title}{Interaction of markov processes}.
\newblock \emph{\bibinfo{journal}{Advances in Mathematics}}
  \textbf{\bibinfo{volume}{5}}, \bibinfo{pages}{246 -- 290}
  (\bibinfo{year}{1970}).
\newblock
  \urlprefix\url{http://www.sciencedirect.com/science/article/pii/0001870870900344}.

\bibitem{93DEM}
\bibinfo{author}{Derrida, B.}, \bibinfo{author}{Evans, M.~R.} \&
  \bibinfo{author}{Mukamel, D.}
\newblock \bibinfo{title}{Exact diffusion constant for one-dimensional
  asymmetric exclusion models}.
\newblock \emph{\bibinfo{journal}{Journal of Physics A: Mathematical and
  General}} \textbf{\bibinfo{volume}{26}}, \bibinfo{pages}{4911}
  (\bibinfo{year}{1993}).
\newblock \urlprefix\url{http://stacks.iop.org/0305-4470/26/i=19/a=023}.

\bibitem{10CCGT}
\bibinfo{author}{Chernyak, V.}, \bibinfo{author}{Chertkov, M.},
  \bibinfo{author}{Goldberg, D.} \& \bibinfo{author}{Turitsyn, K.}
\newblock \bibinfo{title}{Non-equilibrium statistical physics of currents in
  queuing networks}.
\newblock \emph{\bibinfo{journal}{Journal of Statistical Physics}}
  \textbf{\bibinfo{volume}{140}}, \bibinfo{pages}{819--845}
  (\bibinfo{year}{2010}).

\bibitem{01FGV}
\bibinfo{author}{Falkovich, G.}, \bibinfo{author}{Gaw\ifmmode~\mbox{\c{e}}\else
  \c{e}\fi{}dzki, K.} \& \bibinfo{author}{Vergassola, M.}
\newblock \bibinfo{title}{Particles and fields in fluid turbulence}.
\newblock \emph{\bibinfo{journal}{Rev. Mod. Phys.}}
  \textbf{\bibinfo{volume}{73}}, \bibinfo{pages}{913--975}
  (\bibinfo{year}{2001}).
\newblock \urlprefix\url{http://link.aps.org/doi/10.1103/RevModPhys.73.913}.

\end{thebibliography}

\end{document}